# Spontaneous flexo-effect in nanoferroics

Eugene A. Eliseev [*,1], Anna N. Morozovska [†,1], Maya D.Glinchuk[1] and R. Blinc[2],

[2]Jožef Stefan Institute, P. O. Box 3000, 1001 Ljubljana, Slovenia

[1]Institute for Problems of Materials Science, NAS of Ukraine,
Krjijanovskogo 3, 03142 Kiev, Ukraine

**Abstract**

Within the Landau-Ginsburg-Devonshire phenomenological approach we study the ferroic nanosystems properties changes caused by the flexo-effect (flexoelectric, flexomagnetic, flexoelastic) existing spontaneously due to the inhomogeneity of order parameters. Exact solution for the spatially inhomogeneous mechanical displacement vector allowing for flexo-coupling contribution was found for nanowires and thin pills. Strong influence of flexo-effect in nanorods and thin pills leads to the displacements of the atoms resulting into the unit cell symmetry changes, which lead to the phase transition temperature shift, as well as the flat geometry in radial direction transforms into the saucer-like one. The new phenomena can be considered as true manifestation of the spontaneous flexo-effect existence. It was shown that flexo-effect leads to (a) the appearance of new linear and nonlinear contribution and renormalization of coefficients before the order parameter gradient, (b) essentially influences the transition temperature, piezoelectric response and the spatial distribution of the order parameter, (c) results in renormalization of extrapolation length in the boundary conditions. These effects cannot be neglected for ferroelectrics, the renormalization being important for nanoparticles of arbitrary shape, while the linear and nonlinear terms is essential for the thin pills only. They are absent for nanowires with the order parameter directed along the wire axis. We demonstrated that the flexoelectric coupling decreases the polarization gradient self-consistently and so makes polarization more homogeneous. The divergences of dielectric permittivity and correlation radius at some critical value of flexoelectric coefficient originate from the critical radius dependence on the coefficient. This peculiar behavior shows a new way to govern material properties. The effect of the correlation radius renormalization by the flexoelectric effect leads to the renormalization of the intrinsic width of ferroelectric domain walls.

---

[*] Corresponding author: eliseev@i.com.ua

[†] Corresponding author, morozo@i.com.ua; permanent address: V. Lashkarev Institute of Semiconductor Physics, NAS of Ukraine, 41, pr. Nauki, 03028 Kiev, Ukraine

## 1. Introduction

The most general definition of the direct flexo-effect is the appearance of the polarization or magnetization in response to *inhomogeneous* mechanical impact (elastic stress or strain *gradient*). The converse flexo-effect corresponds to the appearance of mechanical stress or strain in response to the *gradient* of order parameter. Let us underline that flexo-coupling affects both the system response to the external impact and the intrinsic gradient of order parameters. Typical example is flexoelectric effect [1, 2, 3] originated from the coupling of polarization gradient with elastic strain and polarization with elastic strain gradient.

It is worth to underline that the flexomagnetic effect [4] should exist in the materials with symmetry without inversion of time (e.g. in some antiferromagnetics), so that the symmetry consideration for flexomagnetic effect will be similar to the well-studied flexoelectric phenomena. In the third primary ferroics ferroelastics the similar effect, called flexoelastic, can be observed in materials without inversion center, where the flexoelastic fifth rank tensor components can be nonzero. It is obvious that flexoelastic effect is absent in cubic symmetry materials, which we consider in the paper.

Up to now the flexoelectric effect was most studied. The phenomenon was firstly predicted by Mashkevich and Tolpygo [5]. Then detailed theoretical study of the flexoelectric effect in bulk crystals was performed by Tagantsev [2, 3] (see also some earlier Refs therein). Experimental measurements of flexoelectric tensor components were carried out by Ma and Cross [6, 7, 8] and Zubko et al. [9]. Renovation of the theoretical description for the flexoelectric response of different nanostructures starts from Catalan et al. papers [10, 11]. Recent achievements are presented in Majdoub et al. [12], Kalinin and Meunier [13] papers. However, in these papers the flexoelectric effect was considered as a coupling between *intrinsic* polar properties (e.g. polarization) and the *extrinsic* factors like the misfit strain relaxation [10, 11] or the system bending by external forces [12, 13], while the coupling between *intrinsic* parameters (i.e. spontaneous polarization gradient and strain) was not considered.

The crucial role of the surface in all physical properties of nanosystems including the strong order parameter gradients in ferroic nanostructures [14] inevitably leads to the noticeable *spontaneous flexo-coupling*, almost negligible for bulk materials, since the order parameters are usually homogeneous in this case.

It is worth to underline that phenomenological approach which was broadly used in the world scientific literature for description of nanosystems [15, 16] as well as in the majority of aforementioned theoretical papers (including the present paper) needs the estimations of sizes range where phenomenological approach is valid. In particular in nanoferroelectrics, where the correlation effects play the decisive role in appearance of the long range order described phenomenologically pretty good the approach can be valid up to the domain wall width, i.e. to the sizes from several to several tens of lattice constants. In more general case the critical sizes of the long-range order existence in ferroics derived on the basis of microscopic [17] and phenomenological calculations [18] are in reasonable agreement both with each other and with experimental observation of ferromagnetic critical size of 7–30 nm [19] and the ferroelectric critical sizes of 1-10 nm [20, 21, 22]. So, keeping in mind that phenomenological approach is suitable for long-range order description, the critical sizes lay in the region from several lattice constants to several nanometers, the region of phenomenological approach validity is broad enough. With respect to flexoelectric effect recently Majdoub et al. [12, 23] considered either cantilever-shaped or thin film nanostructure allowing for flexoelectric effect influence on either response to external stress gradient or polarization gradient on dielectric properties. They apply phenomenological macroscopic models to nanosized systems and compare obtained results with microscopic modeling (*ab initio* and molecular dynamics simulations). They have found that most of the microscopic modeling results are qualitatively (and in some cases quantitatively) reproduced by phenomenological model. Moreover, Majdoub et al. have shown that the inclusion of flexoelectricity into the phenomenological model is essential for quantitative reconciliation of atomistic results for realistic capacitor structure with metallic electrodes.

In the paper we study the ferroic nanostructures properties changes caused by flexo-effect existing due to the inhomogeneous spatial distribution of the order parameter (see Sections 2 and 3). Using concrete example of nanoferroelectrics (mechanically free pills, rods and wires), we demonstrate that the flexoelectric coupling influence the majority of properties and in particular decreases the polarization gradient self-consistently and so renormalizes the correlation radius and stabilizes the ordered phase (see Sections 4).

## 2. Basic equations for flexo-effect contribution in ferroic nanoparticles

For the description of flexo-coupling in ferroic nanoparticles we will use the Landau-Ginsburg-Devonshire (LGD) phenomenological approach [24, 25] with respect to the surface energy, gradient energy, depolarization or demagnetization fields, mechanical stress and flexoeffect.

Since in nanostructures the flexoeffects causes the *internal driving forces* via the order parameter gradients, which perform the virtual work, we need to minimize the Helmholtz free energy $F$ [26, 27, 28, 29, 30]. For ferroics with the second order phase transition corresponding LGD expansion of bulk ($F_V$) and surface ($F_S$) parts of Helmholtz free energy $F$ on the order parameter $\eta$ and strain tensor components $u_{ij}$ have the form:

$$F_V = \int_V d^3r \left( \begin{array}{c} \dfrac{a_{ij}(T)}{2}\eta_i\eta_j + \dfrac{a_{ijkl}}{4}\eta_i\eta_j\eta_k\eta_l - \eta_i\left(E_{0i} + \dfrac{E_i^d}{2}\right) + \dfrac{g_{ijkl}}{2}\left(\dfrac{\partial \eta_i}{\partial x_j}\dfrac{\partial \eta_k}{\partial x_l}\right) \\ -\dfrac{f_{ijkl}}{2}\left(\eta_k\dfrac{\partial u_{ij}}{\partial x_l} - u_{ij}\dfrac{\partial \eta_k}{\partial x_l}\right) - q_{ijkl}u_{ij}\eta_k\eta_l + \dfrac{c_{ijkl}}{2}u_{ij}u_{kl} + \dfrac{v_{ijklmn}}{2}\left(\dfrac{\partial u_{ij}}{\partial x_m}\dfrac{\partial u_{kl}}{\partial x_n}\right) \end{array} \right), \quad (1)$$

$$F_S = \int_S d^2r \left( \dfrac{a_{ij}^S}{2}\eta_i\eta_j + \dfrac{a_{ijkl}^S}{4}\eta_i\eta_j\eta_k\eta_l + \mu_{\alpha\beta}^S u_{\alpha\beta} + d_{ijk}^S u_{ij}\eta_k + \dfrac{w_{ijkl}^S}{2}u_{ij}u_{kl} + ...... \right)$$

Coefficients $a_{ij}(T)$ explicitly depend on temperature $T$. Coefficients $a_{ij}^S$, $a_{ijkl}$, $a_{ijkl}^S$ are supposed to be temperature independent, constants $g_{ijkl}$ and $v_{ijklmn}$ determine magnitude of the gradient energy. Tensors $g_{ijkl}$, $a_{ijkl}$ and $a_{ijkl}^S$ are positively defined. Tensor $w_{jklm}^S$ is the surface excess elastic moduli, $\mu_{\alpha\beta}^S$ is the surface stress tensor [31], [32], $d_{ijk}^S$ is the surface piezoelectric tensor [2, 33]. $q_{ijkl}$ is the bulk striction coefficients; $c_{ijkl}$ are components of elastic stiffness tensor [34].

Tensor $f_{ijkl}$ is the flexo-coupling coefficient tensor [9, 10]. In fact, only the Lifshitz invariant $\dfrac{f_{ijkl}}{2}\left(\eta_k\dfrac{\partial u_{ij}}{\partial x_l} - u_{ij}\dfrac{\partial \eta_k}{\partial x_l}\right)$ is relevant for the bulk contribution. Rigorously speaking, the gradient terms like $v_{ijklmn}(\partial u_{ij}/\partial x_k)(\partial u_{lm}/\partial x_n)$, which was ignored in the Refs. [10-11] for the ferroelectrics, are responsible for the stable smooth distribution of the order parameter at nonzero strain gradients, since the presence of Lifshitz invariant essentially changes the stability conditions [35]. Namely, in the scalar case the inequality $f^2 < gc$ should be valid for the stability of the order parameter smooth distribution. We obtained that in the considered tensorial case the terms $v_{ijklmn}(\partial u_{ij}/\partial x_k)(\partial u_{lm}/\partial x_n)$ can be neglected under the condition $f_{klmn}^2 < g_{ijkl}c_{ijmn}$.

$\mathbf{E}_0$ in Eq.(1) is external field coupled with the order parameter $\mathbf{\eta}$. $\mathbf{E}^d$ is depolarization or demagnetization field that appears due to the nonzero divergence ($div(\mathbf{\eta}) \neq 0$) of order parameter $\mathbf{\eta}$ in confined systems [36, 37].

The equations of state $\delta F_V/\delta \eta_i = 0$ and $\delta F_V/\delta u_{ij} = \sigma_{ij}$ ($\sigma_{jk}$ is the stress tensor, $\delta$ is the symbol of variation derivative) obtained by variation of the bulk free energy (1) should be solved along with the equations of mechanical equilibrium $\partial \sigma_{ij}(\mathbf{x})/\partial x_i = 0$ and compatibility equations equivalent to the mechanical displacement vector $u_i$ continuity [38]. Variation of the surface and bulk free energy (1) on $\eta_i$ yields to Euler-Lagrange equations with the boundary conditions

$$\left. \left( g_{kjim}n_k \dfrac{\partial \eta_m}{\partial x_j} + a_{ij}^S \eta_j + \dfrac{f_{jkim}}{2}u_{jk}n_m \right) \right|_S = 0. \quad (2)$$

Here $n_k$ are components of the external normal to the ferroic surface. The most evident consequence of the flexo-coupling is the inhomogeneous boundary conditions.

In order to demonstrate spontaneus flexo-effect contribution to the nanoferroic properties, hereinafter we neglect the surface excess elastic moduli, surface stress tensor and surface piezoelectric effect contributions into the surface energy (1). We consider *mechanically free* nanoparticles without misfit dislocations, which should lead to the "external" flexo-effect only. The contribution of misfit dislocations into the flexoelectric effect in thin films was considered in details by Catalan et al.

Allowing for essential contribution of elastic strains $u_{ij}$ into the free energy (1), let us firstly proceed with their calculations. Then obtained elastic solution should be substituted into the Euler-Lagrange equations for the order parameter components $\eta_j$.

## 3. Spontaneous flexo-effect influence on elastic fields in nanoferroics.

Hereinafter let us consider a ferroic nanorod with radius $R$, height $h$ and the axially-symmetric one-component order parameter $\eta_3(z,\rho)$ directed along the rod axis $z$ (hereinafter $\rho = \sqrt{x_1^2 + x_2^2}$ and $z = x_3$ are cylindrical coordinates). The external field $\mathbf{E} = (0,0,E_0)$ is also directed along z-axis. Equations of mechanical equilibrium $\partial \sigma_{ij}(\mathbf{x})/\partial x_i = 0$, rewritten for nonzero displacement vector components $u_z(z,\rho)$ and $u_\rho(z,\rho)$, allowing for equation of state $\delta F_V/\delta u_{ij} = c_{ijkl}u_{kl} - q_{ij33}\eta_3^2 + f_{ij3l}\partial \eta_3/\partial x_l = \sigma_{ij}$, acquire the form (see Appendix A):

$$c_{11}\frac{\partial^2 u_z}{\partial z^2} + c_{44}\left(\frac{1}{\rho}\frac{\partial u_z}{\partial \rho} + \frac{\partial^2 u_z}{\partial \rho^2}\right) + (c_{12} + c_{44})\left(\frac{\partial^2 u_\rho}{\partial \rho \partial z} + \frac{1}{\rho}\frac{\partial u_\rho}{\partial z}\right) = 2q_{11}\eta_3\frac{\partial \eta_3}{\partial z} - f_1(\eta_3), \quad (3)$$

$$c_{11}\frac{\partial^2 u_\rho}{\partial \rho^2} + \frac{c_{11}}{\rho}\left(\frac{\partial u_\rho}{\partial \rho} - \frac{u_\rho}{\rho}\right) + c_{44}\frac{\partial^2 u_\rho}{\partial z^2} + (c_{12} + c_{44})\frac{\partial^2 u_z}{\partial \rho \partial z} = 2q_{12}\eta_3\frac{\partial \eta_3}{\partial \rho} - f_2(\eta_3). \quad (4)$$

Here the functions $f_1(\eta_3) = f_{11}\frac{\partial^2 \eta_3}{\partial z^2} + f_{44}\left(\frac{\partial^2 \eta_3}{\partial \rho^2} + \frac{1}{\rho}\frac{\partial \eta_3}{\partial \rho}\right)$ and $f_2(\eta_3) = (f_{12} + f_{44})\frac{\partial^2 \eta_3}{\partial \rho \partial z}$ originated from flexoelectric effect. Voigt notations are used hereinafter. The nanorods are regarded mechanically free, thus corresponding boundary conditions are $\sigma_{z\rho}|_{z=\pm h/2} = \sigma_{zz}|_{z=\pm h/2} = 0$, $\sigma_{z\rho}|_{\rho=R} = 0$ and $\sigma_{\rho\rho}|_{\rho=R} = 0$.

The equation of state $\delta F_V/\delta \eta_3 = 0$ in Voigt notation has the form:

$$\begin{pmatrix} \left( a_1 - 2\left( q_{11}\dfrac{\partial u_z}{\partial z} + q_{12}\left( \dfrac{\partial u_\rho}{\partial \rho} + \dfrac{u_\rho}{\rho} \right) \right) \right)\eta_3 + a_{11}\eta_3^3 \\ - g_{11}\dfrac{\partial^2 \eta_3}{\partial z^2} - g_{12}\left( \dfrac{\partial^2 \eta_3}{\partial \rho^2} + \dfrac{1}{\rho}\dfrac{\partial \eta_3}{\partial \rho} \right) \end{pmatrix} = E_0 + E_3^d + f_3(u). \qquad (5)$$

Here the term $f_3(u) = f_{11}\dfrac{\partial^2 u_z}{\partial z^2} + f_{44}\left( \dfrac{1}{\rho}\dfrac{\partial u_z}{\partial \rho} + \dfrac{\partial^2 u_z}{\partial \rho^2} \right) + (f_{12} + f_{44})\left( \dfrac{\partial^2 u_\rho}{\partial \rho \partial z} + \dfrac{1}{\rho}\dfrac{\partial u_\rho}{\partial z} \right)$ originates from the flexoeffect. Eq.(5) should be supplemented by the boundary conditions (2):

$$\begin{cases} \left( \pm g_{11}\dfrac{\partial \eta_3}{\partial z} + a_1^S \eta_3 \pm \left( \dfrac{f_{12}}{2}\left( \dfrac{u_\rho}{\rho} + \dfrac{\partial u_\rho}{\partial \rho} \right) + \dfrac{f_{11}}{2}\dfrac{\partial u_z}{\partial z} \right) \right)\bigg|_{z=\pm h/2} = 0, \\ \left( g_{12}\dfrac{\partial \eta_3}{\partial \rho} + a_1^S \eta_3 + \dfrac{f_{44}}{2}\left( \dfrac{\partial u_z}{\partial \rho} + \dfrac{\partial u_\rho}{\partial z} \right) \right)\bigg|_{\rho=R} = 0. \end{cases} \qquad (6)$$

(sign "−" for $z=-h/2$, sign "+" for $z=h/2$).

Note that the limiting case $h/R \to 0$ corresponds to the films and thin pills, while the case $R/h \to 0$ corresponds to the wire. Our further analysis will be performed in the two cases of thin pills and wires, films will be considered elsewhere.

The *analytical* solution for mechanical displacements in nanowires (i.e. at $R \ll h$, $\partial \eta_3/\partial z \approx 0$ and $\partial \eta_3/\partial \rho \neq 0$) was derived in Appendix B as:

$$u_\rho = \rho\left( \dfrac{1}{\rho^2}\int_0^\rho \dfrac{q_{12}}{c_{11}}\eta_3^2(\widetilde{\rho})\widetilde{\rho}d\widetilde{\rho} + \dfrac{(c_{11}^2 - c_{11}c_{12} + 2c_{12}^2)q_{12} - 2c_{11}c_{12}q_{11}}{2c_{11}(c_{11} - c_{12})(c_{11} + 2c_{12})}\overline{\eta_3^2} \right), \qquad (7a)$$

$$u_z = z\left( \dfrac{q_{11}(c_{11} + c_{12}) - 2q_{12}c_{12}}{(c_{11} - c_{12})(c_{11} + 2c_{12})}\overline{\eta_3^2} \right) - \dfrac{f_{44}}{c_{44}}\eta_3. \qquad (7b)$$

where $\overline{\eta_3^2} = \dfrac{2}{R^2}\int_0^R \eta_3^2(\widetilde{\rho})\widetilde{\rho}d\widetilde{\rho}$. Note, that in-plane elastic strain $u_{\rho\rho} + u_{\psi\psi} = \dfrac{q_{12}}{c_{11}}\left( \eta_3^2 - \overline{\eta_3^2} \right) + \dfrac{2(c_{11}q_{12} - c_{12}q_{11})}{(c_{11} - c_{12})(c_{11} + 2c_{12})}\overline{\eta_3^2}$ calculated from Eqs.(7) is different from the homogeneous spontaneous strain components in bulk material, which are simply proportional to $\eta_3^2 \equiv \overline{\eta_3^2}$ [36]. Actually, the "intrinsic" term proportional to the difference $\left( \eta_3^2 - \overline{\eta_3^2} \right)$ absent in the bulk solution was omitted in the solution for films used in Refs.[10-11].

The substitution of the shear strain $u_{\rho z} = -\dfrac{f_{44}}{2c_{44}}\dfrac{\partial \eta_3}{\partial \rho}$ calculated from Eqs.(7) into Eq.(5) leads to the gradient coefficient $g_{12}$ renormalization $g_{12}^* = g_{12} - f_{44}^2/c_{44}$ caused by the term $2f_{44}\dfrac{\partial u_{z\rho}}{\partial \rho}$ included into $f_3(u)$. Estimations for ferroelectric PbTiO$_3$ at room temperature gives $f_{44}^2/c_{44} \sim 10^{-11}...10^{-9}$ SI units, which is comparable with typical values of $g_{ij} \sim 10^{-10}$ SI units [39]. So, the renormalization cannot be neglected, but the conditions $g_{ij}^* > 0$ are necessary for the stability of the system without considering higher gradient terms (otherwise the single-domain state becomes unstable even at small $v_{ijklmn}(\partial u_{ij}/\partial x_k)(\partial u_{lm}/\partial x_n)$ values). The perovskite ABO$_3$ lattice deformation caused by spontaneous flexo-effect in nanowires is shown in Fig.1a.

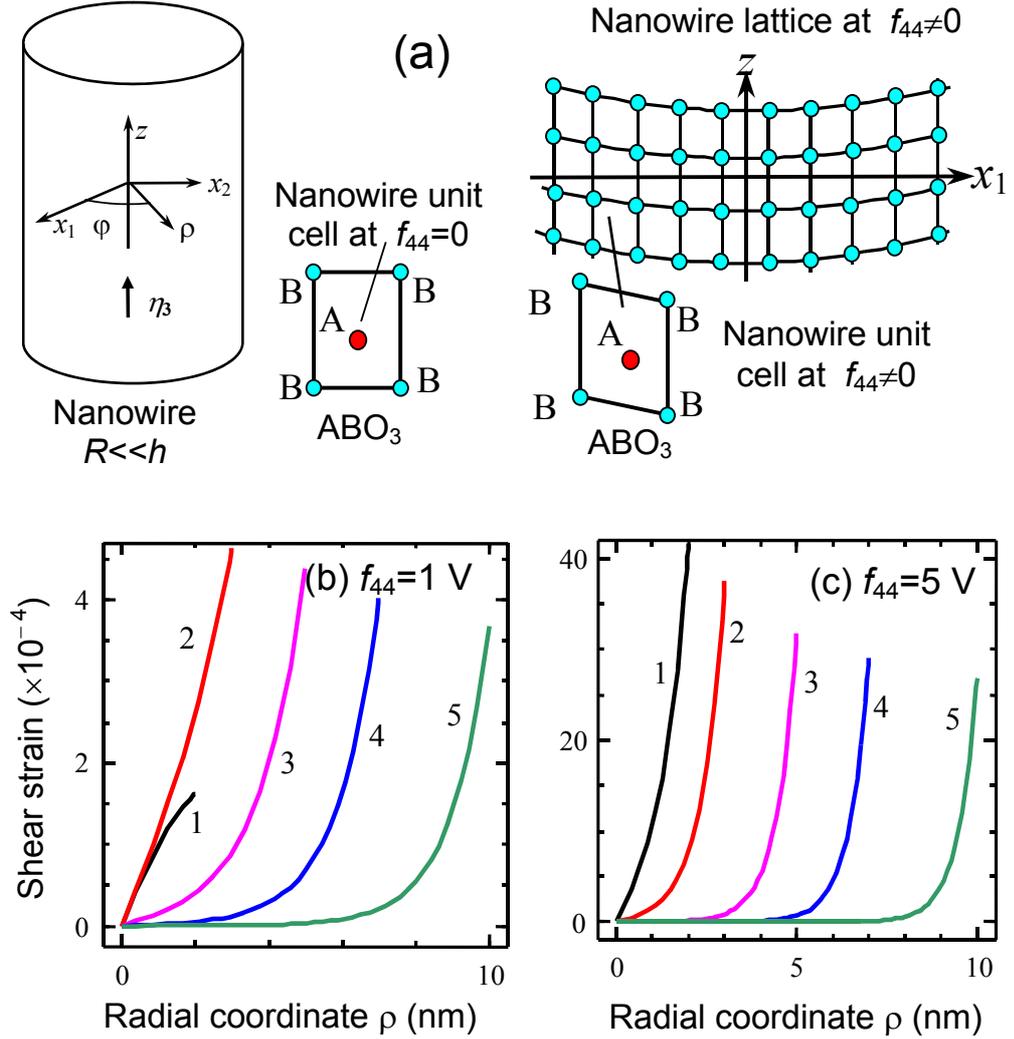

**FIG. 1.** (Color online) (a) Schematics of the perovskite ABO$_3$ lattice radial deformation from squire cross-section to rectangular and incline parallelogram ones caused by spontaneous flexo-effect in nanowires. (b,c) Spontaneous shear strain $u_{\rho z}$ radial distribution inside the nanorod for flexoelectric coefficient $f_{44}=1$ V (b) and 5 V (c). Nanorod radius values are $R = 2, 3, 5, 7, 10$ nm (curves 1-5). PbTiO$_3$ material parameters $T_c = 479^\circ$C, $\alpha_T = 3.8\cdot 10^5$ J m/C$^2$K, $c_{44} = 1.1\cdot 10^{11}$ J/m$^3$, $g_{12} = 10^{-9}$ J m$^3$/C$^2$ and $\lambda_0$=5 nm and $a$=0.4 nm in Eq.(13b).

Typical values of external strain corresponding to the cantilever beam bending [12] are much smaller then the spontaneous one $u_{\rho z} = -(f_{44}/2c_{44})\partial \eta_3/\partial \rho$ caused by spontaneous flexoelectric effect near the surface $\rho = R$ of the nanowire (see Fig.1b,c). Besides the strain influences self-consistently on the order parameter distribution, it should influence on all nanowire electromechanical and electronic properties related with its elastic state. The effect, originated from the inhomogeneity of the order parameter $\partial \eta_3/\partial \rho$, is the stronger the smaller is the wire radius (compare curves 1-6 in Figs.2b,c). Similar effect should exist in thin ferroic nanotubes.

The *analytical* solution for elastic strain and displacements in thin pills (i.e. at $R \gg h$, $\partial \eta_3/\partial z \neq 0$ and $\partial \eta_3/\partial \rho \approx 0$) was derived in Appendix C as

$$u_{zz}(\rho,z) = \left( \frac{q_{11}}{c_{11}} \eta_3^2 + \frac{2c_{12}}{c_{11}} \frac{(c_{12}q_{11} - c_{11}q_{12})\langle \eta_3^2 \rangle}{(c_{11}^2 + c_{11}c_{12} - 2c_{12}^2)} + \frac{12(f_{12}c_{11} - c_{12}f_{11})}{(c_{11}^2 + c_{12}c_{11} - 2c_{12}^2)} \frac{2c_{12}}{c_{11}} \left\langle \frac{\partial \eta_3}{\partial z} \frac{z}{h} \right\rangle \frac{z}{h} - \frac{f_{11}}{c_{11}} \frac{\partial \eta_3}{\partial z} \right), \quad (8a)$$

$$u_z\left(\rho, z = \frac{h}{2}\right) = \frac{12(f_{12}c_{11} - c_{12}f_{11})}{(c_{11}^2 + c_{12}c_{11} - 2c_{12}^2)} \left\langle \frac{\partial \eta_3}{\partial z} \frac{z}{h} \right\rangle \frac{\rho^2}{2h} + h\left( \frac{q_{11}(c_{11}+c_{12}) - 2c_{12}q_{12}}{(c_{11}^2 + c_{11}c_{12} - 2c_{12}^2)} \right)\langle \eta_3^2 \rangle. \quad (8b)$$

Here $\langle \varphi \rangle \equiv \frac{1}{h} \int_{-h/2}^{h/2} \varphi(z)dz$, where $\varphi(z) = \frac{\partial \eta_3}{\partial z} \frac{z}{h}$ or $\eta_3^2(z)$.

Substitution of elastic strain and displacement calculated from Eqs.(8) into Eq.(5) leads to the renormalization of gradient coefficient $g_{11}$ in Eq.(5) $g_{11}^* = g_{11} - f_{11}^2/c_{11}$ due to contribution of the last term $\frac{f_{11}}{c_{11}} \frac{\partial \eta_3}{\partial z}$ in Eq.(8a).

The first term in Eq.(8b) corresponds to the parabolic particle bending induced by the flexo-effect (see Fig. 1a), while the second term is spontaneous strain independent on lateral coordinates (it could be also rewritten as $hQ_{11}\langle \eta_3^2 \rangle$ and typically are much smaller than the first one starting from the radiuses $\rho > 1$ nm). So, the spontaneous flexo-effect leads to the transformation of the flat pill geometry into the saucer-like one. The new phenomenon can be considered as manifestation of spontaneous flexo-effect existence. We hope that it could be observed experimentally.

Allowing for the first term $f_{11} \frac{\partial^2 u_z}{\partial z^2} \equiv f_{11} \frac{\partial u_{zz}}{\partial z}$ in the function $f_3(u)$ introduced in Eq.(5) and expression for out-of-plane strain $u_{zz}$ existing in thin pills, the appearence of *new terms* in the left-hand-side of Eq.(5) was shown. Actually, in particular case of a thin pill ($R>>h$) with symmetric conditions at boundaries $z = \pm h/2$, equation for the order parameter distribution far from the boundary $\rho=R$ can be rewritten as (see Appendix C):

$$\left(a_1 - 4q^2\langle\eta_3^2\rangle + qf\left\langle\frac{\partial\eta_3}{\partial z}\frac{z}{h}\right\rangle\frac{48}{h}z\right)\eta_3 + f^2\left\langle\frac{\partial\eta_3}{\partial z}\frac{z}{h}\right\rangle\frac{24}{h} + b_{11}\eta_3^3 - g_{11}^*\frac{\partial^2\eta_3}{\partial z^2} = E_0 + E_3^d, \quad (9a)$$

where the following designations are introduced:

$$f = \frac{f_{12}c_{11} - c_{12}f_{11}}{\sqrt{c_{11}(c_{11}^2 + c_{11}c_{12} - 2c_{12}^2)}}, \quad q = \frac{q_{12}c_{11} - q_{11}c_{12}}{\sqrt{c_{11}(c_{11}^2 + c_{11}c_{12} - 2c_{12}^2)}}, \quad b_{11} = a_{11} - 2\frac{q_{11}^2}{c_{11}}. \quad (9b)$$

Boundary conditions were obtained from (6) in the form:

$$\left.\left(\lambda^*\frac{\partial\eta_3}{\partial z} \pm \eta_3\right)\right|_{z=\pm h/2} = 0. \quad (9c)$$

Here we introduced the extrapolation length $\lambda^* = \left(g_{11} - \frac{f_{11}^2}{2c_{11}} - f^2\right)\bigg/a_1^S$ renormalized by flexo-effect.

Note, that the new terms proportional to $\frac{f^2}{h}\left\langle\frac{\partial\eta_3}{\partial z}\frac{z}{h}\right\rangle$ and $qf\left\langle\frac{\partial\eta_3}{\partial z}\frac{z}{h}\right\rangle\frac{z}{h}\eta_3$, the latter is related to the flexo-striction coupling contribution, was not considered previously [10-12], since it originates from the intrinsic inhomogeneity of order parameter in the nanostructures. Despite the nonlinear term $\left\langle\frac{\partial\eta_3}{\partial z}\frac{z}{h}\right\rangle\frac{z}{h}\eta_3$ does not lead to the transition temperature shift, our preliminary calculations showed that the term essentially influences (in comparison with conventional cubic term ~ $\eta_3^3$) the spatial distribution of $\eta_3$ in the ordered phase.

The distribution of relative vertical displacement $u_z/h$ caused by flexo-effect for different values of pill thickness $h$ and flexo-coefficient $f_{11}$ is shown in Figs.2b-e for an example of ferroelectric PbTiO$_3$.

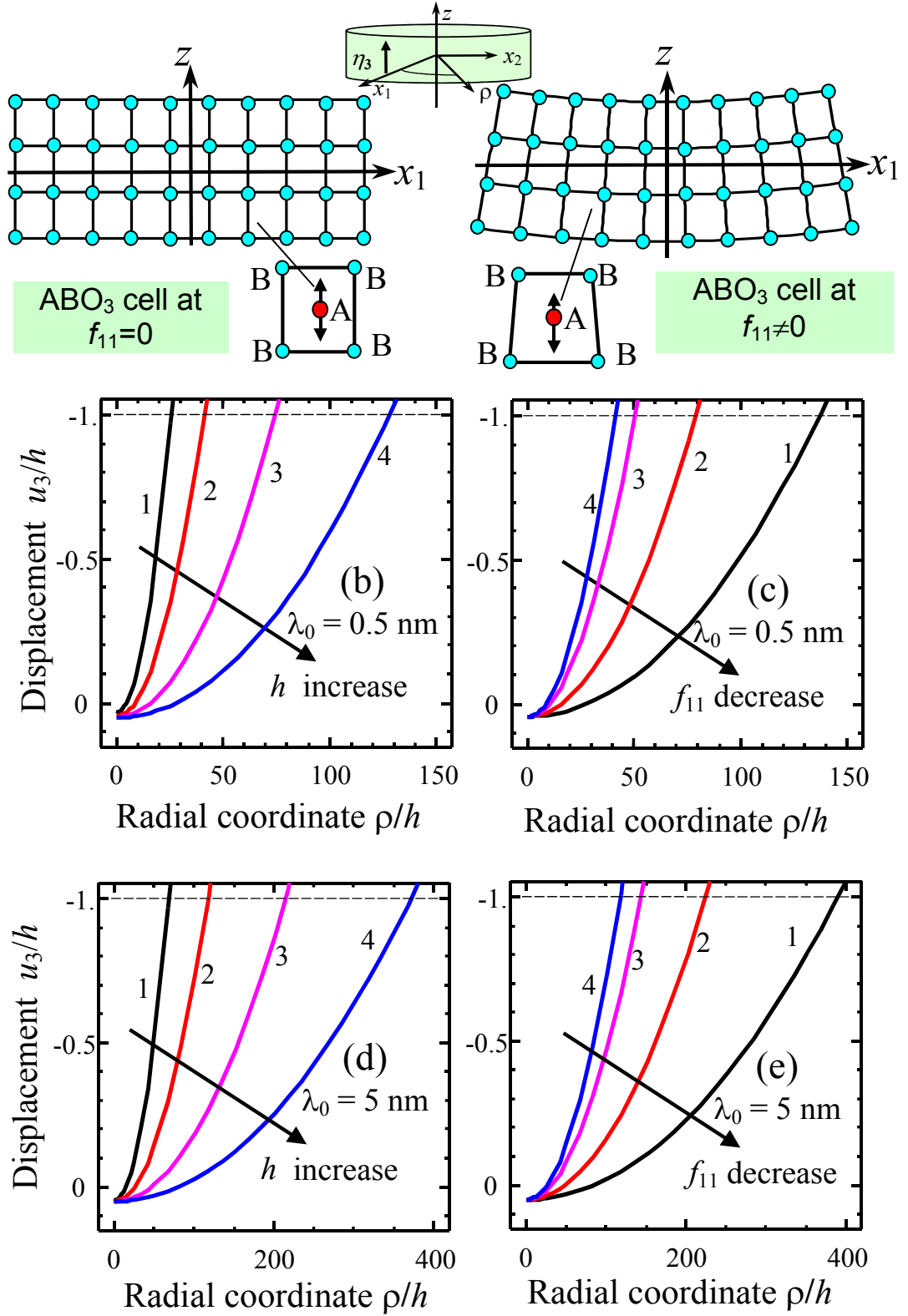

**FIG. 2.** (Color online) (a) Schematics of the perovskite $ABO_3$ lattice deformation caused by spontaneous flexo-effect in nanopills. (b-e) The distribution of vertical displacement for different values of pill thickness $h$=10, 30, 100, 300 nm (curves 1, 2, 3, 4), flexoelectric coefficient for $f_{11}$ = 10 V (b, d); $h$=30 nm, flexoelectric coefficient for $f_{11}$ = 1, 3, 7, 10 V (curves 1, 2, 3, 4) (c, e), seeding extrapolation length $\lambda_0 = g_{11}/a_1^S$ = 0.5 nm (b, c) and 5 nm (d, e), material parameters correspond to $PbTiO_3$.

One can see that relative displacement decreases with pill thickness increase (compare curves 1-4 in parts b,d) and increases with flexoelectric coefficient increase (compare curves 1-4 in parts c,e). The displacement profile is parabolic in agreement with Eq.(8b). Note, that the first flexo-induced term in Eq.(8b) is negative, while the second striction term is positive for PbTiO$_3$ material parameters. For $\rho = 0$ the displacement $u_z\left(0, z = \frac{h}{2}\right) = h\left(\frac{q_{11}(c_{11}+c_{12}) - 2c_{12}q_{12}}{(c_{11}^2 + c_{11}c_{12} - 2c_{12}^2)}\right)\langle\eta_3^2\rangle$ depend on the thickness $h$, but the numerical values of the different curves 1-4 vertical shift appeared ~ 0.05 – 0.03, so it is invisible in the linear scale. Note, that boundary conditions (9c) are valid until $u_3 \ll h$. For small extrapolation lengths displacement increases more rapidly and becomes essential at higher thicknesses $h$ (compare plots a and b, c and d).

In what follows we will consider in more details the influence of flexoelectric effect on the properties of ferroelectric nanos, while the previous calculations are valid for ferroic nanos.

## 4. The influence of spontaneous flexoelectric effect on the properties of nanoferroelectrics

### 4.1 Thin pills

For specificity hereinafter we put $a_1(T) = \alpha_T(T - T_C)$ and use typical for screened ferroelectric thin pills depolarization field $E_3^d = (\langle P_3\rangle - P_3)/(\varepsilon_0\varepsilon_b)$, $P_3$ is polarization directed along the pill symmetry axes (see inset in Fig.2), where $\varepsilon_b$ is dielectric permittivity of the background [40] or reference state [41] unrelated with ferroelectric soft mode (typically $\varepsilon_b<10$). Linearized solution of Eq.(9a) gives the averaged value of susceptibility in paraelectric phase:

$$\langle\chi_{33}\rangle \approx \left(1 - \frac{2R_z^2}{(R_z+\lambda^*)h}\right)\left(a_1 + \frac{2g_{11}^*}{(R_z+\lambda^*)h} - \frac{24f^2 R_z}{(R_z+\lambda^*)h^2}\right)^{-1}. \qquad (10)$$

Here we introduced the characteristic length $R_z^2 = g_{11}^*\varepsilon_0\varepsilon_b$ (see Appendix C for details). Eq.(10) was derived for typical condition $h \gg R_z$. Using the divergence of susceptibility (10), one could find the critical temperature of the transition between paraelectric and ferroelectric phases:

$$T_{cr}(h, f_{11}) \approx T_C - \frac{1}{\alpha_T}\left(\frac{2g_{11}^*}{(R_z+\lambda^*)h} - \frac{24f^2 R_z}{(R_z+\lambda^*)h^2}\right). \qquad (11)$$

The first term in Eq.(11) is the bulk transition temperature, the second term is mainly determined by the influence of surface effects and depolarization field renormalized by the flexoeffect. The third term originated from to the influence of the flexo-term $\sim \frac{f^2}{h}\left\langle\frac{\partial P_3}{\partial z}\frac{z}{h}\right\rangle$ in Eq.(9a). It should be noted that the signs of these terms are different, so, while the second term leads to the critical temperature suppression, the third one always increases the temperature.

The transition temperature non-monotonic behavior, namely minimum at thickness $h_{\min} = 24 f^2 R_z / g_{11}^*$, followed by increase at the smallest thickness values appeared for high values of flexoelectric coefficient $f_{11}$ are related to the third term in Eq.(11) that is inversely proportional to $h^2$. Despite the term is negligible at higher thicknesses, its contribution to the transition temperature dominates over the second term proportional to $1/h$ at small thickness values. However, one should restrict consideration for the thickness greater than several lattice constants, otherwise phenomenological approach can be inapplicable (see Introduction). Since $h_{\min}$ value depends on the material parameters and so it is not excluded that for some materials $h_{\min}$ can be in the region of phenomenological theory applicability.

The ferroelectric transition temperature dependence on thickness $h$ and flexoelectric coefficient $f_{11}$ calculated from exact expression (C.21) is shown in Figs. 3.

However, for the small flexoelectric coefficients values $h_{min}$ is usually smaller than several lattice constants, so the effect of $T_{cr}$ non-monotonic behavior predicted within Landau-Ginzburg-Devonshire phenomenological approach may be unrealistic (see dotted pats of curves 1-4 in the region of ultrasmall thickness in Fig.3a,b). However, for the higher values of flexoelectric coefficient $f_{11}$, and so $h_{min}$ values, the transition temperature $T_{cr}(h, f_{11})$ is rather high at $h=h_{min}$, so the disappearance of thickness-induced phase transition induced by flexoelectric coupling in ferroelectric pills can be reliably predicted within the phenomenology (see curves 3, 4 in Figs.3a,b).

The influence of the extrapolation length on the transition temperature and flexoelectric coupling effect is obvious: for small extrapolation lengths size effects are more pronounced and become essential at higher thicknesses $h$ (compare plots a and b).

The effect of transition temperature increase with flexoelectric coefficient $f_{11}$ increase is demonstrated in Figs.3c,d for several fixed thicknesses $h$. The smaller is the thickness $h$ the higher is the slope of $T_{cr}(f_{11})$ dependence (compare curves 1-4 in Figs.3c,d). Temperature $T_{cr}(f_{11})$ increases with thickness $h$ increase until $h \gg h_{\min}$ (compare curves 4, 3, 2 in Figs.3c,d). For ultra-thin pills with thickness $h \leq h_{\min}$ non-monotonic effects appeared (see maximum at curves 1 in Figs.3c,d). For smaller extrapolation lengths size effects are more pronounced and become essential at higher thicknesses $h$ (compare plots c and d). Note, that all curves in Figs.2c,d have physical meaning until flexoelectric coefficient $f_{11}$ is smaller than the limiting value $f_0 = \sqrt{g_{11} c_{11}}$.

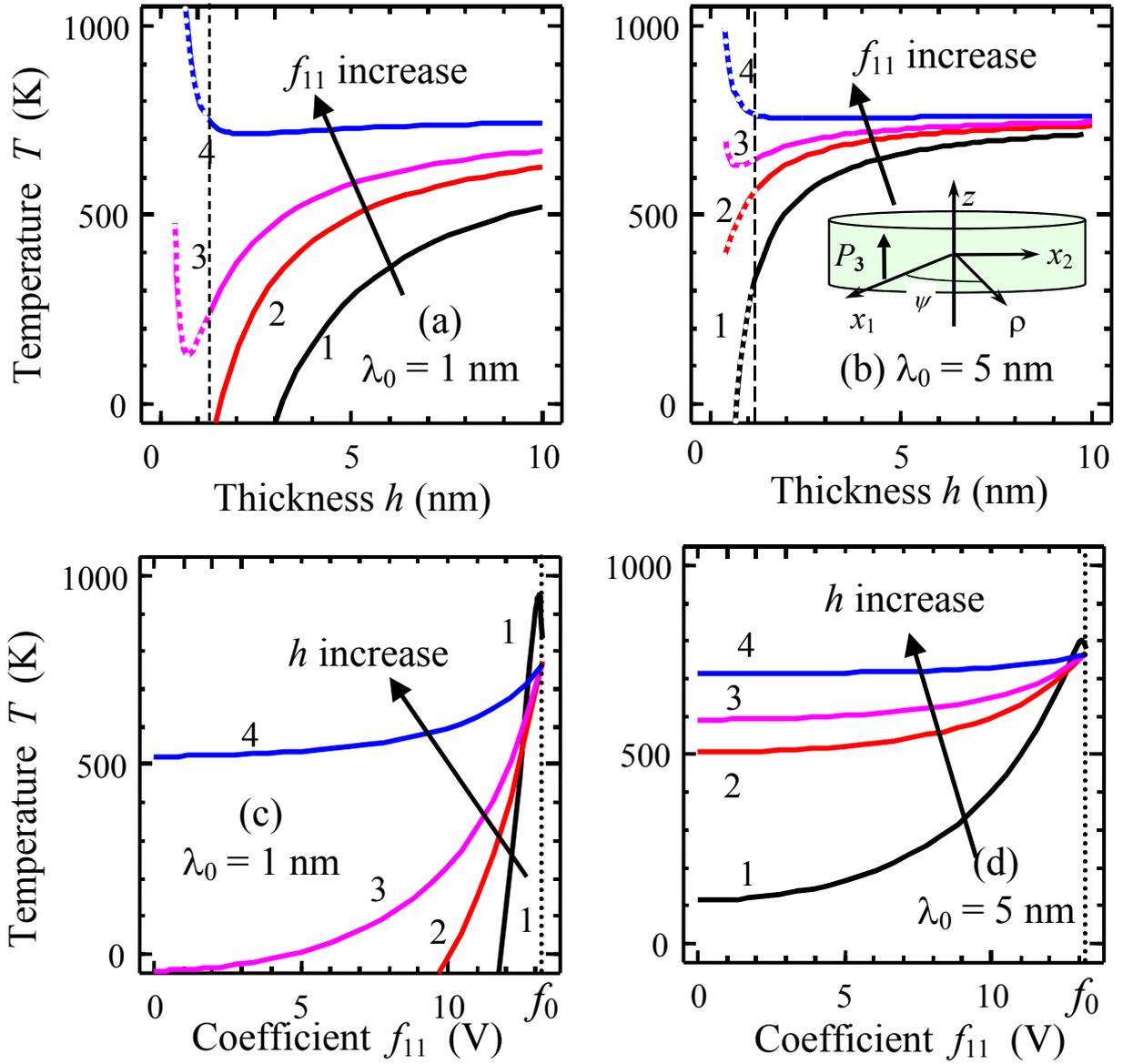

**FIG. 3.** (Color online) (a, b) The dependence of transition temperature $T_{cr}$ on pill thickness $h$ for different values of flexoelectric coefficient $f_{11} = 0, 11, 12, 13$ V (curves 1, 2, 3, 4). (c, d) The dependence of transition temperature $T_{cr}$ on flexoelectric coefficient for different values of thickness $h = 0.8, 2, 3, 10$ nm (curves 1, 2, 3, 4). Seeding extrapolation length $\lambda_0 = g_{11}/a_1^S = 1$ nm (a, c) and 5 nm (b, d), material parameters correspond to PbTiO$_3$: $g_{11} = 10^{-9}$ J m$^3$/C$^2$, $\varepsilon_b = 1$, $T_C = 765$ K, $\alpha_T = 7.53\ 10^6$ J m/C$^2$K, $c_{11} = 1.75\ 10^{11}$ J/m$^3$, $c_{12} = 0.79\ 10^{11}$ J/m$^3$.

Actually, analytical expression Eq.(11) provides quantitative information about the spontaneous flexo-effect contribution into the transition temperature in ferroic nanopills. What is the simple physical meaning of flexo-effect induced increase of the phase transition temperature? The reasonable explanation is the "ordering" role of the new term $\dfrac{f^2}{h}\left\langle \dfrac{\partial P_3}{\partial z}\dfrac{z}{h}\right\rangle$ *linear* on the order parameter $P_3$. The flexo-induced deformation of the ABO$_3$ unit cell vertical cross-section (shown in Fig. 2a) stimulates the polar-active displacement of central B-cation and thus increases paraelectric phase instability. However, linearized solution of Eq.(9a) was obtained in the assumption that the pill surfaces $z = \pm h/2$ remained almost flat (see the boundary conditions (9c)). Rigorously, it is approximation valid until $u_3 \ll h$, since flexo-effect leads to the pill edges bending (as shown in Fig.2b-d).

In the next section we demonstrate that although the terms like $\left\langle \frac{\partial P_3}{\partial z} \frac{z}{h} \right\rangle$ and $\left\langle \frac{\partial P_3}{\partial z} \frac{z}{h} \right\rangle \frac{z}{h} \eta_3$ are absent for ferroelectric nanowire with order parameter directed along its axis flexoelectric effect essentially influences nanowire properties.

*4.2 Nanowires*

Below we demonstrate that the renormalization of gradient coefficient and extrapolation length strongly affects all the properties and in particular the transition temperature shift and correlation radii in single-domain ferroelectric nanowires with axial symmetry of the polarization $P_3$ (see Fig. 3a). For short-circuit boundary conditions $E_3^d(\rho,z) \approx \left(1 + (h/2R)^2\right)^{-1} \left(\overline{P_3} - P_3(\rho,z)\right)/(\varepsilon_0 \varepsilon_b)$ [42], while $E_3^d(\rho,z) \approx -\left(1 + (h/2R)^2\right)^{-1} P_3(\rho,z)/(\varepsilon_0 \varepsilon_b)$ for the open-circuit ones. So, one can neglect small depolarization field $E_d \sim (R/h)^2$ for the case $h \gg R$ typical for nanowires.

Substitution of Eqs.(7) into the Eq.(5) leads to the Euler-Lagrange equation for the polarization $P_3(\rho)$:

$$a_1(T)P_3 - g_{12}^*\left(\frac{d^2 P_3}{d\rho^2} + \frac{1}{\rho}\frac{dP_3}{d\rho}\right) + \left(a_{11} - \frac{q_{12}^2}{c_{11}}\right)P_3^3 + b_{11}\overline{P_3^2}P_3 = E_0, \qquad (12)$$

Renormalized coefficient $b_{11} = -\dfrac{q_{11}^2(c_{11}+c_{12})c_{11} + (c_{11}^2 - c_{11}c_{12} + 2c_{12}^2)q_{12}^2 - 4c_{11}c_{12}q_{11}q_{12}}{c_{11}(c_{11}-c_{12})(c_{11}+2c_{12})}$. The boundary conditions (6) can be rewritten via renormalized extrapolation length $\lambda^*$ as:

$$\left. \left(P_3 + \lambda^*(R)\frac{dP_3}{d\rho}\right)\right|_{\rho=R} = 0, \qquad (13a)$$

$$\lambda^*(R) = \frac{g_{12}}{a_1^S(R)}\left(1 - \frac{f_{44}^2}{2c_{44}g_{12}}\right) = \frac{4R\lambda_0}{4R - 2a + 5\lambda_0}\left(1 - \frac{f_{44}^2}{2c_{44}g_{12}}\right). \qquad (13b)$$

In Eq.(13b) we used that the "seeding" extrapolation length $\lambda(R) = g_{12}/a_1^S(R)$ depends on the rod radius $R$ and material lattice constant $a$ as $\lambda(R) = \dfrac{4R\lambda_0}{4R - 2a + 5\lambda_0}$ in accordance with Wang and Smith calculations [18]. Here $\lambda_0$ has the meaning of extrapolation length of semi-infinite ferroelectric material.

So, for nanowires flexoelectric effect leads to the renormalization of the gradient coefficient and extrapolation length. This means that in order to consider the spontaneous flexoelectric effect influence on the properties one has to rewrite all the analytical expressions obtained earlier for long nanorods and nanowires physical properties without flexoelectric effect [14, 43] by the substitution $g_{12}^*$ and $\lambda^*$ for $g$ and $\lambda_S$ in the expressions for the corresponding property. In what follows we will demonstrate the spontaneous flexo-effect influence on the critical parameters (temperature and radius) of size-induced phase transition and correlation radius using the results [14, 43] obtained without flexoeffect.

Approximate expression for ferroelectric to the paraelectric phase transition temperature $T_{cr}(R)$ for nanowires could be rewritten as:

$$T_{cr}(R, f_{44}) \approx \begin{cases} T_C - \dfrac{2}{\alpha_T}\left(\dfrac{g_{12}^*}{R\lambda^*(R) + 2R^2/k_{01}^2}\right), & \lambda^* > 0, \\ T_C - \dfrac{2}{\alpha_T}\left(g_{12}^* \dfrac{2\lambda^*(R) - R}{2R\lambda^{*2}}\right), & \lambda^* < 0. \end{cases} \quad (14)$$

Where $k_{01} = 2.408...$ is the smallest positive root of equation $J_0(k) = 0$. Renormalized transition temperature $T_{cr}$ dependences vs. nanowire radius and flexoelectric coefficients $f_{44}$ are shown in Figs.4b-c for dimensionless variables, and in Figs.4d-e for PbTiO$_3$ material parameters. It is clear from the plots, that the higher is the $f_{44}$ value, the higher is the transition temperature $T_{cr}$ and the smaller is the critical radius $R_{cr}$ that corresponds to $T = T_{cr}$.

An approximate analytical expression for the critical radius was derived from Eq.(14) under the assumption $R \geq \lambda_0$:

$$R_{cr}(T, f_{44}) \approx \sqrt{\left(1 - \dfrac{f_{44}^2}{f_0^2}\right)k_{01}^2 R_0^2 \dfrac{T_C}{T_C - T} + \left(\dfrac{f_{44}^2}{2f_0^2} - 1\right)^2 k_{01}^4 \dfrac{\lambda^{*2}}{16} - k_{01}^2 \dfrac{\lambda^*}{4}\left(1 - \dfrac{f_{44}^2}{2f_0^2}\right)}. \quad (15)$$

Here $f_0 = \sqrt{g_{12}c_{44}}$, $\lambda^*(R) \approx \lambda_0(1 - f_{44}^2/2c_{44}g_{12})$ and $R_b = \sqrt{g_{12}/\alpha_T T_c}$ is the bulk correlation radius at zero temperature. Under the condition $R < \lambda_0$ the critical radius should be calculated numerically from Eq.(14) as the solution of equation $T_{cr}(R_{cr}, f_{44}) = 0$. Critical radius $R_{cr}$ dependence on flexoelectric coefficients $f_{44}$ for different temperatures is shown in Fig. 5. Solid curves calculated numerically from Eq.(14) are very closed to dashed curves calculated from Eq.(15). So, it is clear that approximation (15) works surprisingly well. Thus, flexoelectric effect renormalizes both critical temperature and critical radius.

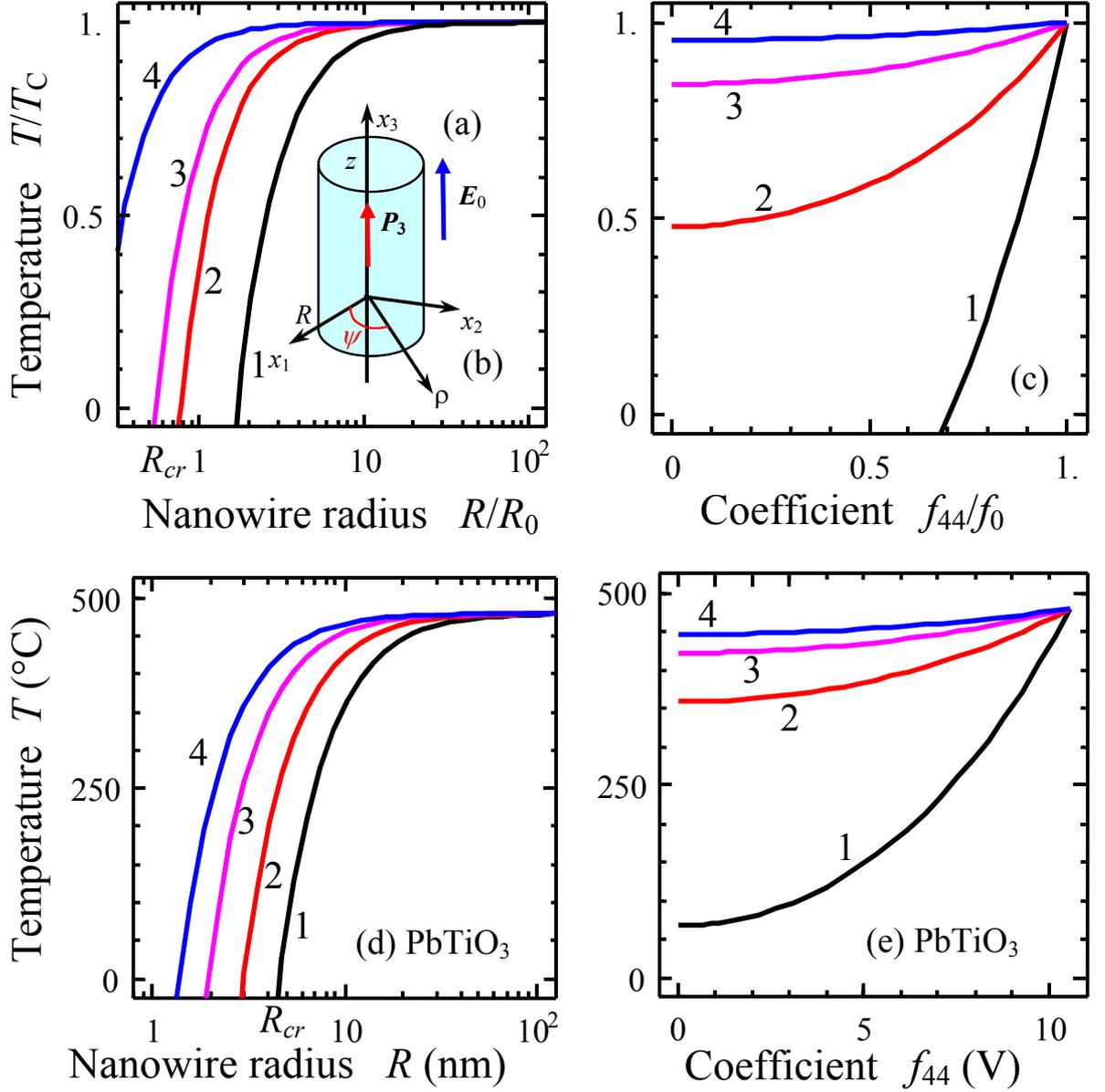

**FIG. 4.** (Color online) (a) Nanowire with cylindrical coordinates defined as $(\rho, \psi, z)$. (b-c) Phase transition temperature $T_{cr}$ dependence vs. the nanowire radius at fixed flexoelectric coefficient $f_{44}/f_0 = 0, 0.9, 0.95, 0.99$ (curves 1, 2, 3, 4) and (c) $T_{cr}$ dependence on flexoelectric coefficient at fixed radius $R/R_c = 1.25, 2.5, 5, 10$ (curves 1, 2, 3, 4). Parameter $g_{12}/(a_1^S R_c) = 1$. (d, e) Transition temperature $T_{cr}$ dependence on radius (d) at fixed values of flexoelectric coefficient $f_{44} = 0, 8, 9.5, 10$ V (curves 1, 2, 3, 4) and dependence on flexoelectric coefficient (e) at fixed values of radius $R = 5, 10, 15, 20$ nm (curves 1, 2, 3, 4). PbTiO$_3$ material parameters $T_c = 479$°C, $\alpha_T = 3.8 \cdot 10^5$ J m/C$^2$K, $c_{44} = 1.1 \cdot 10^{11}$ J/m$^3$, $g_{12} = 10^{-9}$ J m$^3$/C$^2$ and seeding extrapolation length $\lambda_0 = 1$ nm.

Let us underline, that at radiuses slightly higher than the critical one the region of the almost vertical slope of the dependences $T_{cr}(R)$ drastically increases with $f_{44}$ increase (compare curves 1 and 4 in Fig.4b). For chosen material parameters the increase of the slope caused by the flexoelectric coefficient increase on several percents leads to the increase of transition temperature in 2-3 times. With the rod radius increase the terms related with the flexoelectric effect decreases as $1/R$ and becomes unessential at radii $R >> R_{cr}$ (curves 1-4 in Fig.4b,d calculated at different $f_{44}$ converge together and tend to the bulk value with radius increase).

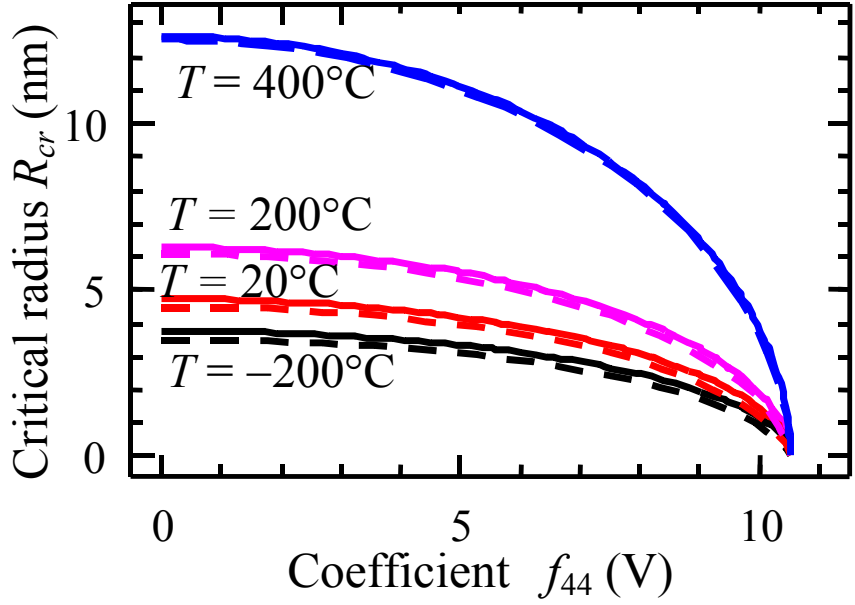

**FIG. 5.** (Color online) Critical radius $R_{cr}$ vs. flexoelectric coefficients $f_{44}$ for different temperatures $T = -20, 20, 200, 400°C$ (marked near the curves). Solid curves are calculated numerically from Eq.(14), dashed curves are calculated from analytical Eq.(15). PbTiO$_3$ material parameters are the same as in Fig.4.

Application of the direct variational method [14, 42] for the Euler-Lagrange Eq.(12) leads to the conventional form of the free energy with renormalized coefficients:

$$F_R(\overline{P}_3) \approx \alpha_T(T - T_{cr}(R))\frac{\overline{P}_3^2}{2} + \left(a_{11} - \frac{q_{12}^2}{c_{11}} + b_{11}\right)\frac{\overline{P}_3^4}{4} - \overline{P}_3 E_0. \tag{16}$$

It is seen, that because the critical temperature $T_{cr}(R)$ in Eq. (14) depends on the flexoelectric coupling coefficient $f_{44}$, the average polarization $\overline{P}_3$ and all other physical properties determined by it have to be dependent on the flexoelectric coefficient $f_{44}$. Note, that $\overline{P}_3$ and other physical properties can be found by the conventional minimization of the free energy (16).

The main effect from the flexoelectric coupling is the change of transition temperature via the renormalization of the extrapolation length and the gradient term (see Eq.(14)). Because of the same reasons flexoelectric coupling will lead to the renormalization of correlation radius as:

$$R_c^*(T,R,f_{44}) = \begin{cases} \sqrt{\dfrac{g_{12} - f_{44}^2/c_{44}}{\alpha_T(T - T_{cr}(R,f_{44}))}}, & \text{paraelectric phase,} \\ \sqrt{-\dfrac{g_{12} - f_{44}^2/c_{44}}{2\alpha_T(T - T_{cr}(R,f_{44}))}}, & \text{ferroelectric phase.} \end{cases} \quad (17)$$

The renormalized correlation radius dependences vs. nanowire radius and flexoelectric coefficients $f_{44}$ are shown in Figs.6a-d for PbTiO$_3$ material parameters at room temperature.

The divergences of correlation radius could be achieved only for $T = T_{cr}(R, f_{44})$ or at $R = R_{cr}$, corresponding to the paraelectric-ferroelectric phase transition point as one can see from Eq.(17). These conditions can be fulfilled at fixed value of radius $R$ for arbitrary value $f_{44}$ or for arbitrary value of radius at given value of temperature $T$, as one can see from Fig. 6. Since the same fixed values of $R$ or $f_{44}$ correspond to the divergence (or maxima for finite electric field value) of dielectric permittivity $\chi$ (because $R_c^* \sim \sqrt{g_{12}^*\chi}$) these values of the radius and flexoelectric coefficient represent the critical radius or "critical" flexoelectric coefficient (corresponding to $R_{cr}$ given by Eq.(15)) of the paraelectric – ferroelectric phase transition.

It is clear from the Figs.6a-b that in ferroelectric phase (i.e. at $R>R_{cr}$) the correlation radius monotonically decreases with the increase of the flexoelectric coefficient $f_{44}$. At the same time, in paraelectric phase correlation radius increases with the increase of the flexoelectric coefficient, since the critical temperature (14) increases with the increase of the flexoelectric coefficient. This opens the possibility to govern the phase diagram and polar properties by the choice of the material with necessary flexoelectric coefficient at given temperature or nanoparticle radius [44].

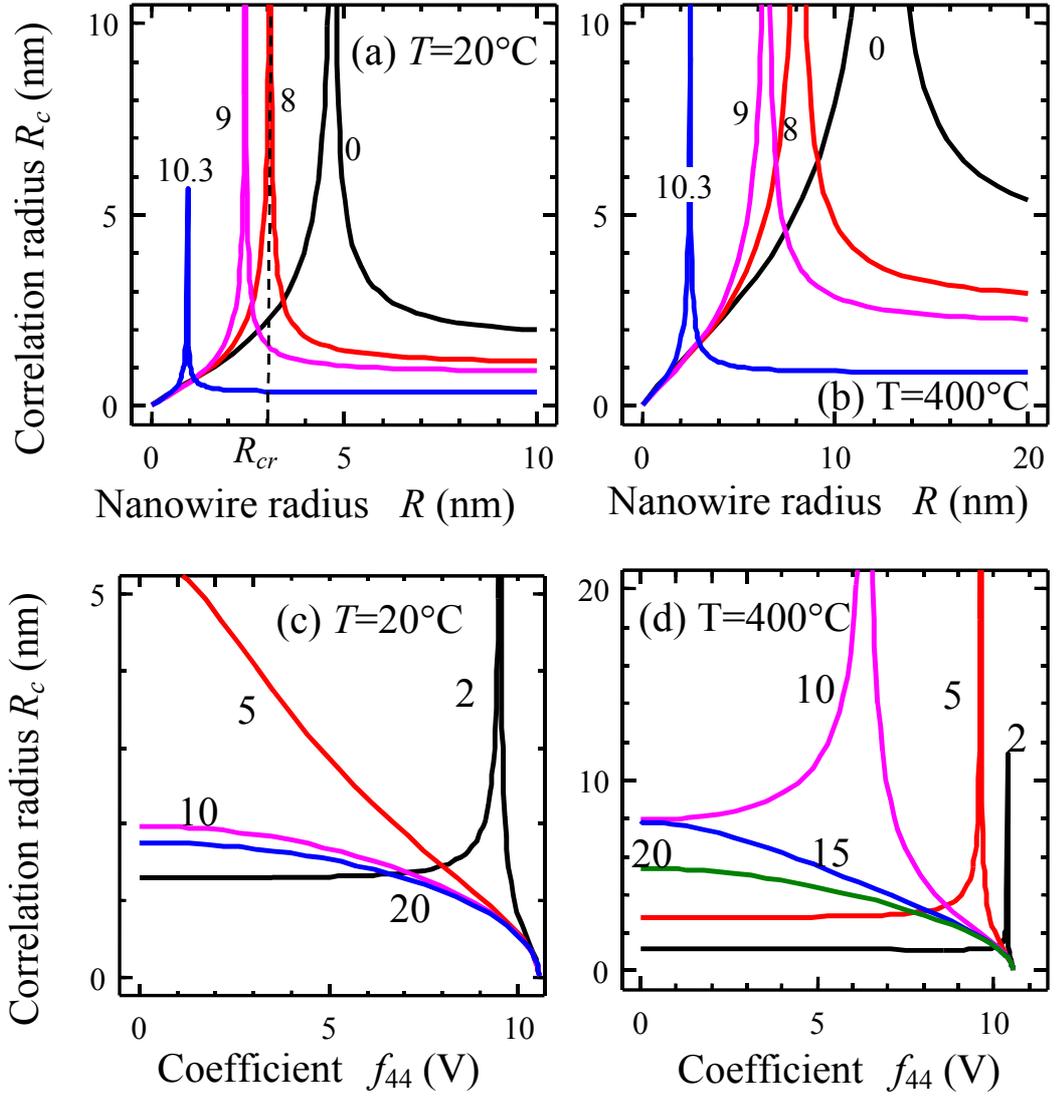

**FIG. 6.** (Color online) (a,b) Correlation radius dependences vs. nanowire radius for different flexoelectric coefficients $f_{44}$ marked near curves (in V) and $T = 20, 400^{\circ}$C. (c,d) Correlation radius dependences vs. flexoelectric coefficients $f_{44}$ for different wire radius marked near curves (in nm) and $T = 20, 400^{\circ}$C. PbTiO$_3$ material parameters are the same as in Fig.4.

Finally, let us calculate the nanowire *piezoelectric reaction* to electric field $E_3$ applied along polar axes z. Using the elastic field (7), one could calculate *piezoelectric reaction* as $d_{kij} = \partial u_{ij}/\partial E_k$. One of the nontrivial consequences of the flexo-effect is the *local appearance* of *new* piezoelectric tensor components, related with the *unit cell deformation* (see Fig.1a), absent in the bulk system:

$$d_{331} = d_{313} = -\frac{f_{44}}{2c_{44}}\frac{\partial \chi_{33}}{\partial \rho}\cos\varphi, \qquad d_{332} = d_{323} = -\frac{f_{44}}{2c_{44}}\frac{\partial \chi_{33}}{\partial \rho}\sin\varphi, \qquad (18)$$

here $\chi_{33}$ is dielectric susceptibility, $\varphi$ is the polar angle. The *flexo-induced* part of the piezoelectric reaction amplitude is proportional to

$$d_{33\rho}(T,R) \sim \frac{f_{44}}{2\alpha_T(T_{cr}(R)-T)}\left(\frac{J_1(\rho/R_0)}{R_0 J_0(R/R_0)-(\lambda^*/R_0)J_1(R/R_0)}\right), \quad (19)$$

where radius $R_0 = \sqrt{g_{12}^*/\alpha_T(T-T_C)}$. As anticipated $d_{33\rho}(T,R)$ diverges in the point $T=T_{cr}(R)$ of size-induced paraelectric-ferroelectric phase transition. It is clear from Figs.7c,d the piezoresponse is the higher the smaller is the wire radius (compare curves 1-6).

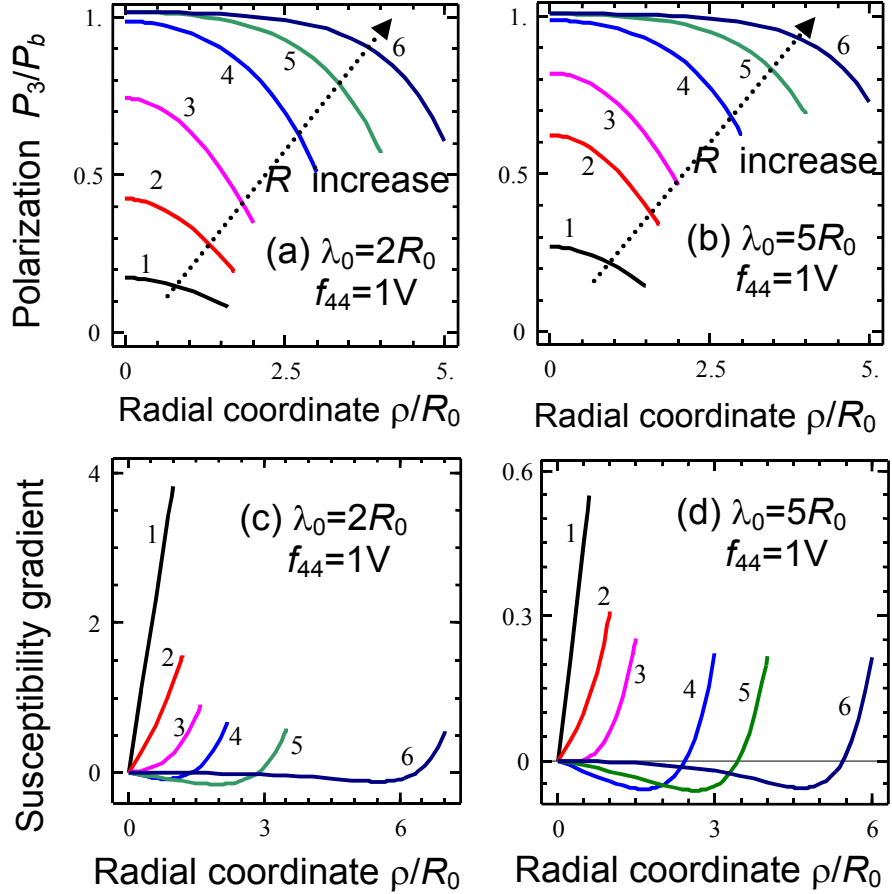

**FIG. 7.** (Color online) (a,b) Polarization distribution inside the nanorods with flexoelectric coefficient $f_{44}=1$ V. Lattice constant $a<0.5R_0$, different values of "seeding" extrapolation length $\lambda_0 =2R_0$ (a) and 5 (b) and nanorod radius values $R/R_0$=1.615, 1.7, 2, 3, 4, 5 (curves 1-6 in plot (a)) and $R/R_0$=1.5, 1.7, 2, 3, 4, 5 (curves 1-6 in plot (b)). Here $P_b = \sqrt{-\alpha_T(T-T_c)/a_{11}}$ is spontaneous polarization of the bulk material. (c,d) Normalized susceptibility gradient $(\partial\chi_{33}/\partial\rho)\chi_b^{-1}R_0$ inside the nanorod of different radius $R$ for flexoelectric coefficient $f_{44}=1$ V and PbTiO$_3$ material parameters. Lattice constant $a<0.5R_0$, different values of "seeding" extrapolation length $\lambda_0 =2R_0$ (a) and 5 (b) and nanorod radius values $R/R_0$=1, 1.2, 1.6, 2.2, 3.5, 7, 10 (curves 1-6 in plot (a)) and $R/R_0$=0.6, 1, 1.5, 3, 4, 6 (curves 1-6 in plot (b)).

Note, that the flexo-contribution (19) is proportional to $(T_{cr}(R)-T)^{-1}$, so it can be much greater then the striction contribution $d_{3ij} \sim -2q_{ij33}P_3\chi_{33}$ proportional to the $(T_{cr}(R)-T)^{-1/2}$ near the size-induced phase transition point $T = T_{cr}(R)$. Thus, spontaneous flexo-effect originated from the intrinsic gradient $\partial P_3/\partial \rho$ of the order parameter near the wire surface could lead to *giant piezoelectric reaction.* It should be noted that measured coefficient of proportionality between strain gradient and polarization is $\chi_{33} f_{ij3l}$.[8]

**5. Discussion**

The inhomogeneity of order parameters being inevitably present in all ferroics nanosystems and originating from the surface influence is the source of spontaneous flexo-effect (flexoelectric and flexomagnetic) coupling with mechanical strain. Thus, the spontaneous flexo-effect is unavoidable because the inhomogeneity of order parameters in all nanosystems is direct consequence of surface effects, thus it has to be taken into account when calculating properties especially for comparison of the theory with experiment.

The exact analytical solution for the spatially inhomogeneous mechanical displacement vector allowing for flexo-effect contribution was derived for nanowires and pills with one-component vector order parameter $\eta_3$ directed along the nanoparticle symmetry axis. We show that flexo-effect leads to appearance of two additional terms in the equation of state for the order parameter $\eta_3$, proportional to its first derivative, and the third term proportional to its second derivative. These terms respectively cause the appearance of new linear term $\sim \langle x_3 \partial \eta_3/\partial x_3 \rangle$ and nonlinear contribution $\sim \eta_3 \langle x_3 \partial \eta_3/\partial x_3 \rangle$ in thin pills, the renormalization of coefficients before the order parameter gradient $\partial^2 \eta_3/\partial x_i \partial x_j$, as well as result in inhomogeneity and renormalization of extrapolation length in the boundary conditions for pills and nanowires. Estimations show that these effects cannot be neglected.

The spontaneous flexo-effect leads to the transformation of the unit cell symmetry (e.g. from the squire cross-section to trapezoid one) of rods and pills that changes the flat geometry in radial direction into the saucer-like one. The new phenomenon can be considered as manifestation of spontaneous flexo-effect existence. The forecast is waiting for experimental verification.

The influence of flexo-effect on the nanosystem properties was considered in details for the most studied flexoelectric effect. One can conclude that even a rather moderate flexoelectric effect significantly renormalizes all the polar, piezoelectric and dielectric properties and in particular the correlation radius, suppresses the size-induced phase transition from ferroelectric to paraelectric phase and thus stabilizes the ordered phase in ferroic nanoparticles.

The divergences of dielectric permittivity and correlation radius at the "critical" value of the flexoelectric coefficient related to the critical radius had shown a new way to govern ferroelectric materials properties. The effect of the correlation radius renormalization by the flexoelectric effect leads to the changes of the domain wall intrinsic width. The predicted effects are useful for design of ferroelectric nanowires with radius up to several nanometers, which have ultra-thin domain walls and reveal close to bulk polar properties.

**Appendix A. Free energy functional and elastic problem for flexoelectric ferroelectric nanoparticles**

For perovskite (cubic) symmetry the free energy expansion (1) in powers of polarization $P_3$ and strain $u_{nm}$ has the form:

$$F_V = \int_V d^3r \left( \begin{array}{c} \dfrac{a_1(T)}{2}P_3^2 + \dfrac{a_{11}}{4}P_3^4 + \dfrac{g}{2}(\nabla P_3)^2 - P_3\left(E_0 + \dfrac{E_3^d}{2}\right) - q_{ij33}u_{ij}P_3^2 \\ -\dfrac{f_{ij3l}}{2}\left(P_3 \dfrac{\partial u_{ij}}{\partial x_l} - u_{ij}\dfrac{\partial P_3}{\partial x_l}\right) + \dfrac{c_{ijkl}}{2}u_{ij}u_{kl} + \dfrac{v_{ijkl}}{2}\left(\nabla_i \nabla_j u_{kl}\right)^2 \end{array} \right). \quad (A.1)$$

$$F_S = \int_S d^2r \left( \dfrac{a_1^S}{2}P_3^2 + \dfrac{a_{11}^S}{4}P_3^4 + ... + \dfrac{w_{jklm}^S}{2}u_{jk}u_{lm} \right)$$

Free energy (5) is minimal when polarization $P_3$ and relevant strain or stress tensor components are defined at the nanostructure boundaries [45]. Under such conditions, one should solve equations of state (5):

$$\left(a_1 - 2q_{ij33}u_{ij}\right)P_3 + a_{11}P_3^3 - f_{ij3l}\dfrac{\partial u_{ij}}{\partial x_l} - g\dfrac{\partial^2 P_3}{\partial x_k \partial x_k} = E_0 + E_3^d, \quad (A.2a)$$

$$f_{ij3l}\dfrac{\partial P_3}{\partial x_l} - q_{ij33}P_3^2 + c_{ijkl}u_{kl} - v_{ijkl}\dfrac{\partial^2 u_{kl}}{\partial x_m \partial x_m} = \sigma_{ij}. \quad (A.2b)$$

Eqs.(A.2) should be supplemented by the boundary conditions for strain (or stress) and polarization. To the best of our knowledge the general solution of the coupled problem given by Eqs.(A.2) is absent.

Eqs.(A.2) should be supplemented by the boundary conditions (2), (4) for polarization and strain:

$$\left.\left(g_{ij33}n_i \dfrac{\partial P_3}{\partial x_j} + a_1^S P_3 + \dfrac{f_{ij3m}}{2}u_{ij}n_m\right)\right|_S = 0, \quad \left.\left(v_{ijklmp}n_k \dfrac{\partial u_{lm}}{\partial x_p} + w_{ijkl}^S u_{kl} - \dfrac{f_{ij3m}}{2}P_3 n_m\right)\right|_S = 0, \quad (A.3)$$

For the scalar case the inequality $a(T)v > f^2 - gc$ should be valid in high temperature paraelectric or paramagnetic phrase ($a(T) > 0$), while the single-domain state stability in ordered low temperature ferroelectric or ferromagnetic phase is possible under the condition $-2a(T)v > f^2 - gc$. So, for the considered case the terms $v_{ijklmn}\left(\partial u_{ij}/\partial x_k\right)\left(\partial u_{lm}/\partial x_n\right)$ can be neglected under the condition $f_{klmn}^2 < g_{ijkl}c_{ijmn}$ only.

**Appendix B. Elastic problem solution**

For cubic $O_h$ symmetry materials the flexoelectric tensor has only 3 independent components: $f_{1111}=f_{2222}=f_{3333}\equiv f_{11}$, $f_{1122}=f_{1133}=f_{2211}=f_{2233}=f_{3311}=f_{3322}\equiv f_{12}$ and $f_{1221}=f_{1313}=f_{2323}\equiv f_{44}$. The elastic stiffness modules are $c_{1111}=c_{2222}=c_{3333}\equiv c_{11}$, $c_{1122}=c_{1133}=c_{2211}=c_{2233}=c_{3311}=c_{3322}\equiv c_{12}$ and $c_{1221}=c_{1313}=c_{2323}\equiv c_{44}$. Similar relations holds for elastic compliance modules $s_{1111}=s_{2222}=s_{3333}\equiv s_{11}$, $s_{1122}=s_{1133}=s_{2211}=s_{2233}=s_{3311}=s_{3322}\equiv s_{12}$ and $s_{1221}=s_{1313}=s_{2323}\equiv 4s_{44}$ (note the factor "4" inserted to keep relations like $\sigma=cu$ and $u=s\sigma$ the same form both in Cartesian and matrix notations).

In what follows we will use a *perturbation approach* in the form of the *decoupling approximation* for mechanical and electrostatic equations allowing for the boundary conditions on the nanostructure surfaces.

Minimization of free energy on elastic stresses, $\partial F_V/\partial u_{ij}=\sigma_{ij}$, gives the following set of equations of state:

$$\begin{cases} \sigma_{11} = c_{11}u_{11} + c_{12}(u_{22}+u_{33}) - q_{12}P_3^2 + f_{12}\frac{\partial P_3}{\partial x_3}, \\ \sigma_{22} = c_{11}u_{22} + c_{12}(u_{11}+u_{33}) - q_{12}P_3^2 + f_{12}\frac{\partial P_3}{\partial x_3}, \\ \sigma_{33} = c_{11}u_{33} + c_{12}(u_{22}+u_{11}) - q_{11}P_3^2 + f_{11}\frac{\partial P_3}{\partial x_3}, \\ \sigma_{23} = 2c_{44}u_{23} + f_{44}\frac{\partial P_3}{\partial x_2}, \quad \sigma_{13} = 2c_{44}u_{13} + f_{44}\frac{\partial P_3}{\partial x_1}, \quad \sigma_{12} = 2c_{44}u_{12}. \end{cases} \quad (B.1)$$

These equations must be supplemented by the equilibrium conditions of bulk and surface forces, namely, $\partial\sigma_{ij}/\partial x_j=0$ in the bulk and $\sigma_{ij}n_j|_S=0$ at the free surface of the system [46]. Note, that for some cases these conditions should be not applied in the points of non-deformed body. Instead one should take into account that points of forces may shift from their initial positions. Thus, "S" denotes the surface of a real system, deformed by the forces.

From general symmetry consideration we can suggest that the displacement vector has only the radial, $u_\rho(z,\rho)$, and axial, $u_z(z,\rho)$, components, which in turns depend only on $\rho$ and $z$ coordinates. In this case components of strain tensor in cylindrical coordinate system are:

$$u_{zz}(z,\rho) = \frac{\partial u_z(z,\rho)}{\partial z}, \quad u_{z\rho}(z,\rho) = \frac{1}{2}\left(\frac{\partial u_\rho(z,\rho)}{\partial z} + \frac{\partial u_z(z,\rho)}{\partial \rho}\right),$$
$$u_{\rho\rho}(z,\rho) = \frac{\partial u_\rho(z,\rho)}{\partial \rho}, \quad u_{\psi\psi}(z,\rho) = \frac{u_\rho(z,\rho)}{\rho}, \quad u_{\rho\psi} = 0, \quad u_{z\psi} = 0. \quad (B.2)$$

The bulk equilibrium conditions in cylindrical coordinates $(\rho,\psi,z)$ are

$$\begin{cases} \dfrac{\partial \sigma_{zz}}{\partial z} + \dfrac{\partial \sigma_{z\rho}}{\partial \rho} + \dfrac{\sigma_{z\rho}}{\rho} + \dfrac{1}{\rho}\dfrac{\partial \sigma_{z\psi}}{\partial \psi} = 0, \\ \dfrac{\partial \sigma_{\rho\rho}}{\partial \rho} + \dfrac{\sigma_{\rho\rho} - \sigma_{\psi\psi}}{\rho} + \dfrac{1}{\rho}\dfrac{\partial \sigma_{\rho\psi}}{\partial \psi} + \dfrac{\partial \sigma_{z\rho}}{\partial z} = 0, \\ \dfrac{1}{\rho}\dfrac{\partial \sigma_{\psi\psi}}{\partial \psi} + \dfrac{\partial \sigma_{z\psi}}{\partial z} + \dfrac{\partial \sigma_{\rho\psi}}{\partial \rho} + 2\dfrac{\sigma_{\rho\psi}}{\rho} = 0. \end{cases} \qquad (B.3)$$

Rewriting the relevant equations of state (B.1) via the stresses gives the following:

$$\begin{cases} \sigma_{\rho\rho} = c_{11}u_{\rho\rho} + c_{12}(u_{\psi\psi} + u_{zz}) - q_{12}P_3^2 + f_{12}\dfrac{\partial P_3}{\partial z}, \\ \sigma_{\psi\psi} = c_{11}u_{\psi\psi} + c_{12}(u_{\rho\rho} + u_{zz}) - q_{12}P_3^2 + f_{12}\dfrac{\partial P_3}{\partial z}, \\ \sigma_{zz} = c_{11}u_{zz} + c_{12}(u_{\psi\psi} + u_{\rho\rho}) - q_{11}P_3^2 + f_{11}\dfrac{\partial P_3}{\partial z}, \\ \sigma_{z\rho} = 2c_{44}u_{z\rho} + f_{44}\dfrac{\partial P_3}{\partial \rho}, \quad \sigma_{z\psi} = 2c_{44}u_{z\psi}, \quad \sigma_{\rho\psi} = 2c_{44}u_{\rho\psi}. \end{cases} \qquad (B.4a)$$

It is obvious, that $\sigma_{z\psi} = \sigma_{\rho\psi} = 0$, since $u_{z\psi} = u_{\rho\psi} = 0$. Next we rewrite the equations of state (B.4a) via displacement components allowing for Eqs. (B.2) as:

$$\begin{cases} \sigma_{\rho\rho} = c_{11}\dfrac{\partial u_\rho}{\partial \rho} + c_{12}\left(\dfrac{u_\rho}{\rho} + \dfrac{\partial u_z}{\partial z}\right) - q_{12}P_3^2 + f_{12}\dfrac{\partial P_3}{\partial z}, \\ \sigma_{\psi\psi} = c_{11}\dfrac{u_\rho}{\rho} + c_{12}\left(\dfrac{\partial u_\rho}{\partial \rho} + \dfrac{\partial u_z}{\partial z}\right) - q_{12}P_3^2 + f_{12}\dfrac{\partial P_3}{\partial z}, \\ \sigma_{zz} = c_{11}\dfrac{\partial u_z}{\partial z} + c_{12}\left(\dfrac{u_\rho}{\rho} + \dfrac{\partial u_\rho}{\partial \rho}\right) - q_{11}P_3^2 + f_{11}\dfrac{\partial P_3}{\partial z}, \\ \sigma_{z\rho} = c_{44}\left(\dfrac{\partial u_\rho}{\partial z} + \dfrac{\partial u_z}{\partial \rho}\right) + f_{44}\dfrac{\partial P_3}{\partial \rho}. \end{cases} \qquad (B.4b)$$

In the considered case the equilibrium conditions (B.3) in bulk reduce to

$$\begin{cases} \dfrac{\partial \sigma_{zz}}{\partial z} + \dfrac{\partial \sigma_{z\rho}}{\partial \rho} + \dfrac{\sigma_{z\rho}}{\rho} = 0, \\ \dfrac{\partial \sigma_{\rho\rho}}{\partial \rho} + \dfrac{\sigma_{\rho\rho} - \sigma_{\psi\psi}}{\rho} + \dfrac{\partial \sigma_{z\rho}}{\partial z} = 0. \end{cases} \qquad (B.5)$$

Using the equations of state in the form (B.4b), it is easy to rewrite (B.5) as

$$\begin{cases} \dfrac{\partial}{\partial z}\left(c_{11}u_{zz}+c_{12}(u_{\psi\psi}+u_{\rho\rho})-q_{11}P_3^2+f_{11}\dfrac{\partial P_3}{\partial z}\right)+ \\ +\dfrac{\partial}{\partial\rho}\left(2c_{44}u_{z\rho}+f_{44}\dfrac{\partial P_3}{\partial\rho}\right)+\dfrac{1}{\rho}\left(2c_{44}u_{z\rho}+f_{44}\dfrac{\partial P_3}{\partial\rho}\right)=0, \\ \dfrac{\partial}{\partial\rho}\left(c_{11}u_{\rho\rho}+c_{12}(u_{\psi\psi}+u_{zz})-q_{12}P_3^2+f_{12}\dfrac{\partial P_3}{\partial z}\right)+ \\ +\dfrac{c_{11}-c_{12}}{\rho}(u_{\rho\rho}-u_{\psi\psi})+\dfrac{\partial}{\partial z}\left(2c_{44}u_{z\rho}+f_{44}\dfrac{\partial P_3}{\partial\rho}\right)=0. \end{cases}$$

Using definition of strain via the mechanical displacement vector, Eq. (B.2), we have

$$\begin{pmatrix} c_{11}\dfrac{\partial^2 u_z}{\partial z^2}+c_{44}\dfrac{\partial^2 u_z}{\partial\rho^2}+\dfrac{c_{44}}{\rho}\dfrac{\partial u_z}{\partial\rho} \\ +(c_{12}+c_{44})\left(\dfrac{\partial^2 u_\rho}{\partial\rho\partial z}+\dfrac{1}{\rho}\dfrac{\partial u_\rho}{\partial z}\right) \end{pmatrix}=q_{11}\dfrac{\partial}{\partial z}P_3^2-f_{11}\dfrac{\partial^2 P_3}{\partial z^2}-f_{44}\dfrac{\partial^2 P_3}{\partial\rho^2}-\dfrac{f_{44}}{\rho}\dfrac{\partial P_3}{\partial\rho},\quad\text{(B.6a)}$$

$$\begin{pmatrix} c_{11}\dfrac{\partial^2 u_\rho}{\partial\rho^2}+\dfrac{c_{11}}{\rho}\left(\dfrac{\partial u_\rho}{\partial\rho}-\dfrac{u_\rho}{\rho}\right)+c_{44}\dfrac{\partial^2 u_\rho}{\partial z^2} \\ +(c_{12}+c_{44})\dfrac{\partial^2 u_z}{\partial\rho\partial z} \end{pmatrix}=q_{12}\dfrac{\partial}{\partial\rho}P_3^2-f_{12}\dfrac{\partial^2 P_3}{\partial\rho\partial z}-f_{44}\dfrac{\partial^2 P_3}{\partial\rho\partial z}.\quad\text{(B.6b)}$$

These equations should be supplemented by the conditions of absence of normal stress components at rod surface, $n_k\sigma_{kj}=0$, namely at side surface $\rho=R$:

$$\sigma_{\rho\rho}=\left(c_{11}\dfrac{\partial u_\rho}{\partial\rho}+c_{12}\left(\dfrac{u_\rho}{\rho}+\dfrac{\partial u_z}{\partial z}\right)-q_{12}P_3^2+f_{12}\dfrac{\partial P_3}{\partial z}\right)\bigg|_{\rho=R}=0 \quad\text{(B.7a)}$$

$$\sigma_{z\rho}=\left(c_{44}\left(\dfrac{\partial u_\rho}{\partial z}+\dfrac{\partial u_z}{\partial\rho}\right)+f_{44}\dfrac{\partial P_3}{\partial\rho}\right)\bigg|_{\rho=R}=0.\quad\text{(B.7b)}$$

and rod faces, $z=0, h$:

$$\sigma_{z\rho}=\left(c_{44}\left(\dfrac{\partial u_\rho}{\partial z}+\dfrac{\partial u_z}{\partial\rho}\right)+f_{44}\dfrac{\partial P_3}{\partial\rho}\right)\bigg|_{z=0,h}=0,\quad\text{(B.7c)}$$

$$\sigma_{zz}=\left(c_{11}\dfrac{\partial u_z}{\partial z}+c_{12}\left(\dfrac{u_\rho}{\rho}+\dfrac{\partial u_\rho}{\partial\rho}\right)-q_{11}P_3^2+f_{11}\dfrac{\partial P_3}{\partial z}\right)\bigg|_{z=0,h}=0.\quad\text{(B.7d)}$$

For an infinite nanowire when all the observable physical quantities depends only on $\rho$, the system of equations (B.6a) and (B.6b) could be decoupled into two separate equations for $u_z$ and $u_\rho$ respectively. The equation (B.6a) in this case has a trivial solution, $u_z=az-f_{44}P_3/c_{44}$, where the first and second terms determine the dilatational and shear strains respectively. The equation (B.6b) in this case is:

$$\frac{\partial^2 u_\rho}{\partial \rho^2} + \frac{1}{\rho}\frac{\partial u_\rho}{\partial \rho} - \frac{u_\rho}{\rho^2} = \frac{\partial}{\partial \rho}F[P_3], \tag{B.8}$$

$$F[P_3] = \frac{q_{12}}{c_{11}}P_3^2 - \frac{f_{12}}{c_{11}}\frac{\partial P_3}{\partial z} \tag{B.9}$$

The solution of Eqs.(B.6)-(B.8) gives the full displacement and strain field:

$$\begin{cases} u_{zz} = a, \quad u_\rho = \frac{1}{\rho}\int_0^\rho F[P_3(\tilde\rho)]\tilde\rho d\tilde\rho + b\rho + \frac{c}{\rho}, \\ u_{\rho\rho} = F[P_3(\rho)] - \frac{1}{\rho^2}\int_0^\rho F[P_3(\tilde\rho)]\tilde\rho d\tilde\rho + b - \frac{c}{\rho^2}, \\ u_{\psi\psi} = \frac{1}{\rho^2}\int_0^\rho F[P_3(\tilde\rho)]\tilde\rho d\tilde\rho + b + \frac{c}{\rho^2} \\ u_{\rho z} = -\frac{f_{44}}{2c_{44}}\frac{\partial P_3}{\partial \rho}. \end{cases} \tag{B.10}$$

Here we also took into account that Eq. (B.6a) has a trivial solution $u_{zz} = a = const$.
Substitution of (B.10) into (B.4) gives the following expressions for the stress field.

$$\sigma_{\rho\rho} = a c_{12} + b(c_{11} + c_{12}) - \frac{1}{\rho^2}\left(\int_0^\rho F[P_3(\tilde\rho)]\tilde\rho d\tilde\rho + c\right)(c_{11} - c_{12}), \quad \sigma_{\rho z} = 0 \tag{B.11a}$$

$$\sigma_{\psi\psi} = \begin{pmatrix} a c_{12} + b(c_{11} + c_{12}) + \frac{1}{\rho^2}\left(\int_0^\rho F[P_3(\tilde\rho)]\tilde\rho d\tilde\rho + c\right)(c_{11} - c_{12}) \\ + \left(\frac{c_{12}}{c_{11}} - 1\right)\left(q_{12}P_3^2 - 2f_{12}\frac{\partial P_3}{\partial z}\right) \end{pmatrix} \tag{B.11b}$$

$$\sigma_{zz} = a c_{11} + 2b c_{12} + \left(\frac{c_{12}}{c_{11}}q_{12} - q_{11}\right)P_3^2 - \left(\frac{c_{12}}{c_{11}}f_{12} - f_{11}\right)\frac{\partial P_3}{\partial z} \tag{B.11c}$$

Constants $a$, $b$ and $c$ could be found from the boundary conditions.

In the case of freestanding rod stresses should be finite at $\rho \to 0$. Since integral terms are finite for finite $P_3(\rho \to 0)$, this condition means $c=0$, thus (B.11) could be rewritten as

$$\sigma_{\rho\rho} = a c_{12} + b(c_{11} + c_{12}) - \frac{1}{\rho^2}\int_0^\rho F[P_3(\tilde\rho)]\tilde\rho d\tilde\rho (c_{11} - c_{12}) \tag{B.12a}$$

$$\sigma_{\psi\psi} = a c_{12} + b(c_{11} + c_{12}) + \left(\frac{1}{\rho^2}\int_0^\rho F[P_3(\tilde\rho)]\tilde\rho d\tilde\rho - F[P_3(\rho)]\right)(c_{11} - c_{12}), \tag{B.12b}$$

$$\sigma_{zz} = a c_{11} + 2b c_{12} + \left(\frac{c_{12}}{c_{11}}q_{12} - q_{11}\right)P_3^2 - \left(\frac{c_{12}}{c_{11}}f_{12} - f_{11}\right)\frac{\partial P_3}{\partial z} \tag{B.12c}$$

Boundary conditions for the free nanorod surfaces are

$$\sigma_{\rho\rho}(\rho = R) = 0 \tag{B.13a}$$

$$\sigma_{zz}(z = h) = 0 \tag{B.13b}$$

However, as one can see from (B.10c), for the case of arbitrary radial distribution of $P_3$ the condition (B.13b) could not be satisfied in every point of the rod faces. It is the "price" of above made approximation. One of the methods to built physically relevant solution is to fulfill the condition (B.10c) in the sense of Saint-Venant principle, namely one should set to zero full force acting on the free face. Thus, one can get the following system of equations for the coefficients $a$ and $b$:

$$a c_{12} + b(c_{11} + c_{12}) = \left( \frac{q_{12}}{c_{11}} \overline{P_3^2} - 2 \frac{f_{12}}{c_{11}} \overline{\frac{\partial P_3}{\partial z}} \right) \frac{(c_{11} - c_{12})}{2} \tag{B.14a}$$

$$a c_{11} + 2b c_{12} = -\left( \frac{c_{12}}{c_{11}} q_{12} - q_{11} \right) \overline{P_3^2} + 2\left( \frac{c_{12}}{c_{11}} f_{12} - f_{11} \right) \overline{\frac{\partial P_3}{\partial z}} \tag{B.14a}$$

Here $\overline{P_3^2} = \dfrac{2}{R^2} \int_0^R P_3^2(\tilde{\rho}) \tilde{\rho} d\tilde{\rho}$ is the mean polarization of the rod.

Neglecting the terms $\partial \overline{P_3}/\partial z$, one can find the solution of system (B.12) in the form:

$$a = \frac{q_{11}(c_{11} + c_{12}) - 2q_{12}c_{12}}{(c_{11} - c_{12})(c_{11} + 2c_{12})} \overline{P_3^2} \tag{B.15a}$$

$$b = \frac{(c_{11}^2 - c_{11}c_{12} + 2c_{12}^2)q_{12} - 2c_{11}c_{12}q_{11}}{2c_{11}(c_{11} - c_{12})(c_{11} + 2c_{12})} \overline{P_3^2} \tag{B.15b}$$

Now, using Eqs. (B.10) and (B.13), one can write the strain components

$$\begin{aligned}
u_{zz} &= \frac{q_{11}(c_{11} + c_{12}) - 2q_{12}c_{12}}{(c_{11} - c_{12})(c_{11} + 2c_{12})} \overline{P_3^2} \\
u_{\rho\rho} &= \frac{q_{12}}{c_{11}} P_3^2 - \frac{1}{\rho^2} \int_0^\rho \frac{q_{12}}{c_{11}} P_3^2(\tilde{\rho}) \tilde{\rho} d\tilde{\rho} + \frac{(c_{11}^2 - c_{11}c_{12} + 2c_{12}^2)q_{12} - 2c_{11}c_{12}q_{11}}{2c_{11}(c_{11} - c_{12})(c_{11} + 2c_{12})} \overline{P_3^2}, \\
u_{\rho z} &= -\frac{f_{44}}{2c_{44}} \frac{\partial P_3}{\partial \rho}, \\
u_{\psi\psi} &= \frac{1}{\rho^2} \int_0^\rho \frac{q_{12}}{c_{11}} P_3^2(\tilde{\rho}) \tilde{\rho} d\tilde{\rho} + \frac{(c_{11}^2 - c_{11}c_{12} + 2c_{12}^2)q_{12} - 2c_{11}c_{12}q_{11}}{2c_{11}(c_{11} - c_{12})(c_{11} + 2c_{12})} \overline{P_3^2}.
\end{aligned} \tag{B.16}$$

Finally, in Cartesian coordinates the strain tensor components are:

$$u_{\rho\rho} + u_{\psi\psi} = u_{11} + u_{22} = \frac{q_{12}}{c_{11}} P_3^2 + \frac{(c_{11}^2 - c_{11}c_{12} + 2c_{12}^2)q_{12} - 2c_{11}c_{12}q_{11}}{c_{11}(c_{11} - c_{12})(c_{11} + 2c_{12})} \overline{P_3^2}, \tag{B.17a}$$

$$u_{33} = \frac{q_{11}(c_{11} + c_{12}) - 2q_{12}c_{12}}{(c_{11} - c_{12})(c_{11} + 2c_{12})} \overline{P_3^2},$$

$$u_{23} = -\frac{f_{44}}{2c_{44}} \frac{\partial P_3}{\partial x_2}, \quad u_{13} = -\frac{f_{44}}{2c_{44}} \frac{\partial P_3}{\partial x_1}, \quad u_{12} = (u_{\rho\rho} - u_{\psi\psi})\cos\psi\sin\psi$$

(B.17b)

**Appendix C. Flexoelectricity and size effect of ferroelectric thin films**

Equations of state could be found by the variation of the free energy (1), $\delta F/\delta P_3 = 0$ and $\delta F/\delta u_{ij} = \sigma_{ij}$, which gives the following:

$$(a_1 - 2q_{ij33}u_{ij})\eta_3 + a_{11}\eta_3^3 - f_{ij3l}\frac{\partial u_{ij}}{\partial x_l} - g_{ij33}\frac{\partial^2 \eta_3}{\partial x_k \partial x_k} = E_0 + E_3^d. \quad (C.1)$$

$$c_{ijkl}u_{kl} - q_{ij33}\eta_3^2 + f_{ij3l}\frac{\partial \eta_3}{\partial x_l} = \sigma_{ij}, \quad (C.2)$$

where $\sigma_{ij}$ is the strain tensor. Eqs.(C.1) should be supplemented by the boundary conditions for polarization and elastic stresses:

$$\left( g_{ij33}n_i \frac{\partial \eta_3}{\partial x_j} + a_1^S \eta_3 + \frac{f_{ij3m}}{2}u_{ij}n_m \right)\bigg|_S = 0 \quad (C.3)$$

while elastic stress should satisfy the equation of state (C.2) along with conditions of mechanical equilibrium in the bulk

$$\frac{\partial \sigma_{ij}}{\partial x_i} = 0. \quad (C.4)$$

and at the free surfaces

$$\sigma_{ij}n_j\big|_S = 0 \quad (C.5)$$

Considering one-dimensional distributions it is more convenient to find strain and stress fields directly, without switching to displacement components. In this case equations (C.4) and (C.5) reduces to $\partial \sigma_{i3}/\partial x_3 = 0$ and $\sigma_{i3}\big|_S = 0$ (with $i$=1, 2, 3). Thus, components $\sigma_{i3}$ are zero throughout the sample. Thus, for uniaxial ferroelectric only some of diagonal components of strain and stress tensors differ from zero. From the symmetry consideration it is obvious that $\sigma_{11}=\sigma_{22}\equiv\sigma$ and $u_{11}=u_{22}\equiv u$.

Next one could recall the conditions of elastic compatibility $\mathrm{inc}(i,j,\hat{u}) = e_{ikl}e_{jmn}u_{ln,km} = 0$.[38] For the considered case they could be reduced to $\partial^2 u/\partial x_3^2=0$ (while the distribution of $u_{33}$ can be arbitrary along $x_3$). Obvious solution is

$$u = u_0 + u_x \frac{x_3}{h} \quad (C.6)$$

Constants $u_0$ and $u_x$ should be found from boundary conditions for either stress component σ for free films or strain for clamped films. For the first case two conditions should be met

$$\langle\sigma(x_3)\rangle \equiv \frac{1}{h}\int_{-h/2}^{h/2}\sigma(x_3)dx_3 = 0, \quad \left\langle\sigma(x_3)\frac{x_3}{h}\right\rangle \equiv \frac{1}{h}\int_{-h/2}^{h/2}\sigma(x_3)\frac{x_3}{h}dx_3 = 0, \tag{C.7}$$

which means zero total force and total moment acting on the pill.

Introducing matrix notations as follows $c_{11}\equiv c_{1111}$, $c_{12}\equiv c_{1122}$, $q_{11}\equiv q_{1111}$, $q_{12}\equiv q_{1122}$, $f_{11}\equiv f_{1111}$, $f_{12}\equiv f_{1122}$, it is easy to rewrite strain, stress and equation for polarization as

Using these relations, it is easy to rewrite Eqs. (C.1) and (C.2) as follows

$$(a_1 - 4q_{12}u - 2q_{11}u_{33})\eta_3 + a_{11}\eta_3^3 - 2f_{12}\frac{\partial u}{\partial x_3} - f_{11}\frac{\partial u_{33}}{\partial x_3} - g_{11}\frac{\partial^2 \eta_3}{\partial x_3^2} = E_0 + E_3^d. \tag{C.8}$$

$$(c_{11} + c_{12})u + c_{12}u_{33} - q_{12}\eta_3^2 + f_{12}\frac{\partial \eta_3}{\partial x_3} = \sigma, \tag{C.9}$$

$$2c_{12}u + c_{11}u_{33} - q_{11}\eta_3^2 + f_{11}\frac{\partial \eta_3}{\partial x_3} = 0 \tag{C.10}$$

It is easy to find $u_{33}$ component from Eq. (C.10) as

$$u_{33} = \frac{q_{11}}{c_{11}}\eta_3^2 - \frac{f_{11}}{c_{11}}\frac{\partial \eta_3}{\partial x_3} - \frac{2c_{12}}{c_{11}}u \tag{C.11}$$

After substitution of (C.11) into Eqs. (C.8), (C.9) one can get the following

$$\left(a_1 - 4\left(q_{12} - q_{11}\frac{c_{12}}{c_{11}}\right)u\right)\eta_3 + \left(a_{11} - 2\frac{q_{11}^2}{c_{11}}\right)\eta_3^3 - 2\left(f_{12} - f_{11}\frac{c_{12}}{c_{11}}\right)\frac{\partial u}{\partial x_3} - \left(g_{11} - \frac{f_{11}^2}{c_{11}}\right)\frac{\partial^2 \eta_3}{\partial x_3^2} = E_0 + E_3^d \tag{C.12}$$

$$(c_{11} + c_{12})u + c_{12}\left(\frac{q_{11}}{c_{11}}\eta_3^2 - \frac{f_{11}}{c_{11}}\frac{\partial \eta_3}{\partial x_3} - \frac{2c_{12}}{c_{11}}u\right) - q_{12}\eta_3^2 + f_{12}\frac{\partial \eta_3}{\partial x_3} = \sigma \tag{C.13a}$$

Now we could recall solution (C.6) for strain component $u$ and rewrite (C.13a) as

$$\left(c_{11} + c_{12} - \frac{2c_{12}^2}{c_{11}}\right)\left(u_0 + u_x\frac{x_3}{h}\right) + \left(\frac{c_{12}}{c_{11}}q_{11} - q_{12}\right)\eta_3^2 + \left(f_{12} - \frac{c_{12}}{c_{11}}f_{11}\right)\frac{\partial \eta_3}{\partial x_3} = \sigma \tag{C.13b}$$

In order to find two constants, $u_0$ and $u_x$, one should use conditions (C.7) for in-plain stress:

$$\langle\sigma\rangle \equiv \left(c_{11} + c_{12} - \frac{2c_{12}^2}{c_{11}}\right)u_0 + \left(\frac{c_{12}}{c_{11}}q_{11} - q_{12}\right)\langle\eta_3^2\rangle + \left(f_{12} - \frac{c_{12}}{c_{11}}f_{11}\right)\left\langle\frac{\partial \eta_3}{\partial x_3}\right\rangle = 0 \tag{C.13c}$$

$$\left\langle\sigma\frac{x_3}{H}\right\rangle = \left(c_{11} + c_{12} - \frac{2c_{12}^2}{c_{11}}\right)\left(u_x\left\langle\left(\frac{x_3}{h}\right)^2\right\rangle\right) + \left(\frac{c_{12}}{c_{11}}q_{11} - q_{12}\right)\left\langle\eta_3^2\frac{x_3}{h}\right\rangle + \left(f_{12} - \frac{c_{12}}{c_{11}}f_{11}\right)\left\langle\frac{\partial \eta_3}{\partial x_3}\frac{x_3}{h}\right\rangle = 0$$

$$\tag{C.13d}$$

Taking into account the symmetry of polarization distribution, $\eta_3(-x_3) = \eta_3(x_3)$, the solution of the system (C.13c)-(C.13d) with respect to $u_0$ and $u_x$ is

$$u_0 = -\frac{(c_{12}q_{11} - c_{11}q_{12})\langle \eta_3^2 \rangle}{(c_{11}^2 + c_{11}c_{12} - 2c_{12}^2)} \qquad (C.14)$$

$$u_x = -\frac{12(f_{12}c_{11} - c_{12}f_{11})}{(c_{11}^2 + c_{12}c_{11} - 2c_{12}^2)}\left\langle \frac{\partial \eta_3}{\partial x_3}\frac{x_3}{h} \right\rangle \qquad (C.15)$$

Finally, one could find strain components from Eqs. (C.6), (C.11), (C.14) and (C.15) in the form:

$$u = -\frac{(c_{12}q_{11} - c_{11}q_{12})\langle \eta_3^2 \rangle}{(c_{11}^2 + c_{11}c_{12} - 2c_{12}^2)} - \frac{12(f_{12}c_{11} - c_{12}f_{11})}{(c_{11}^2 + c_{12}c_{11} - 2c_{12}^2)}\left\langle \frac{\partial \eta_3}{\partial x_3}\frac{x_3}{h} \right\rangle \frac{x_3}{h}, \qquad (C.16)$$

$$u_{33} = \frac{q_{11}}{c_{11}}\eta_3^2 + \frac{2c_{12}}{c_{11}}\frac{(c_{12}q_{11} - c_{11}q_{12})\langle \eta_3^2 \rangle}{(c_{11}^2 + c_{11}c_{12} - 2c_{12}^2)} + \frac{12(f_{12}c_{11} - c_{12}f_{11})}{(c_{11}^2 + c_{12}c_{11} - 2c_{12}^2)}\frac{2c_{12}}{c_{11}}\left\langle \frac{\partial \eta_3}{\partial x_3}\frac{x_3}{h} \right\rangle \frac{x_3}{h} - \frac{f_{11}}{c_{11}}\frac{\partial \eta_3}{\partial x_3}. \qquad (C.17)$$

Finally, using Eqs. (C.12) and (C.16), one could get the equation, determining the polarization distribution:

$$a_1\eta_3 + \frac{2(f_{12}c_{11} - c_{12}f_{11})^2}{c_{11}(c_{11}^2 + c_{12}c_{11} - 2c_{12}^2)}\left\langle \frac{\partial \eta_3}{\partial x_3}\frac{x_3}{h} \right\rangle \frac{12}{h} + \left(a_{11} - 2\frac{q_{11}^2}{c_{11}}\right)\eta_3^3 - \frac{4(c_{12}q_{11} - c_{11}q_{12})^2\langle \eta_3^2 \rangle \eta_3}{c_{11}(c_{11}^2 + c_{11}c_{12} - 2c_{12}^2)} +$$

$$+ 4\frac{(q_{12}c_{11} - q_{11}c_{12})(f_{12}c_{11} - c_{12}f_{11})}{c_{11}(c_{11}^2 + c_{12}c_{11} - 2c_{12}^2)}\left\langle \frac{\partial \eta_3}{\partial x_3}\frac{x_3}{h} \right\rangle 12\frac{x_3}{h}\eta_3 - \left(g_{11} - \frac{f_{11}^2}{c_{11}}\right)\frac{\partial^2 \eta_3}{\partial x_3^2} = E_0 + E_3^d$$

(C.18)

Boundary conditions for polarization (C.3) could be written as (here we neglected surface flexoelectric and piezoelectric effects):

$$\left(g_{11}n_3\frac{\partial \eta_3}{\partial x_3} + a_1^S \eta_3 + \left(f_{12}u + \frac{f_{11}}{2}u_{33}\right)n_3\right)\bigg|_S = 0 \qquad (C.19a)$$

Using strain field (C.16), (C.17) and neglecting nonlinear terms, one could rewrite (C.19a) as

$$\left(n_3\left(g_{11} - \frac{f_{11}^2}{2c_{11}}\right)\frac{\partial \eta_3}{\partial x_3} - \frac{(f_{12}c_{11} - c_{12}f_{11})^2}{c_{11}(c_{11}^2 + c_{12}c_{11} - 2c_{12}^2)}\left\langle \frac{\partial \eta_3}{\partial x_3}\frac{x_3}{h} \right\rangle\frac{x_3}{h}12n_3 + a_1^S\eta_3\right)\bigg|_S = 0 \qquad (C.19b)$$

Using Simpson's rule, it is easy to evaluate average gradient as

$$\left\langle \frac{\partial \eta_3}{\partial x_3}\frac{x_3}{h} \right\rangle \approx \frac{1}{12}\left(-\frac{\partial \eta_3}{\partial x_3}\bigg|_{x_3 \to -\frac{h}{2}} + \frac{\partial \eta_3}{\partial x_3}\bigg|_{x_3 \to \frac{h}{2}}\right) = \frac{1}{6}\frac{\partial \eta_3}{\partial x_3}\bigg|_{x_3 \to \frac{h}{2}}$$

and obtain finally the boundary condition in evident form:

$$\left(\left(g_{11} - \frac{f_{11}^2}{2c_{11}} - \frac{(f_{12}c_{11} - c_{12}f_{11})^2}{c_{11}(c_{11}^2 + c_{12}c_{11} - 2c_{12}^2)}\right)\frac{\partial \eta_3}{\partial x_3} + a_1^S\eta_3\right)\bigg|_{x_3 \to \frac{h}{2}} = 0 \qquad (C.19c)$$

Now we could introduce the extrapolation length renormalized by flexo-effect

$$\lambda^* = \left(g_{11} - \frac{f_{11}^2}{2c_{11}} - \frac{(f_{12}c_{11} - c_{12}f_{11})^2}{c_{11}(c_{11}^2 + c_{12}c_{11} - 2c_{12}^2)}\right) \bigg/ a_1^S . \tag{C.20}$$

Using typical expression for the depolarization field in ambient charges screened ferroelectric thin pills $E_3^d = (\langle\eta_3\rangle - \eta_3)/(\varepsilon_0\varepsilon_b)$, the linearized solution of (C.18) could be found by standard technique. After the simple, but cumbersome manipulations, one could find average value of order parameter as

$$\langle\eta_3\rangle \left( a_1 + \frac{\frac{2R_z}{h}\sinh\left(\frac{h}{2R_z}\right)\frac{1}{\varepsilon_0\varepsilon_b}}{\cosh\left(\frac{h}{2R_z}\right) + \frac{\lambda^*}{R_z}\sinh\left(\frac{h}{2R_z}\right)} - \frac{2(f_{12}c_{11} - c_{12}f_{11})^2}{c_{11}(c_{11}^2 + c_{12}c_{11} - 2c_{12}^2)}\frac{12}{h^2}\frac{\cosh\left(\frac{h}{2R_z}\right) - \frac{2R_z}{h}\sinh\left(\frac{h}{2R_z}\right)}{\cosh\left(\frac{h}{2R_z}\right) + \frac{\lambda^*}{R_z}\sinh\left(\frac{h}{2R_z}\right)} \right) =$$

$$= E_0\left(1 - \frac{\frac{2R_z}{h}\sinh\left(\frac{h}{2R_z}\right)}{\cosh\left(\frac{h}{2R_z}\right) + \frac{\lambda^*}{R_z}\sinh\left(\frac{h}{2R_z}\right)}\right)$$

(C.21)

Here the new designation $R_z^2 \approx \left(g_{11} - \frac{f_{11}^2}{c_{11}}\right)\varepsilon_0\varepsilon_b$ is introduced. At the limit $h \gg R_z$ Eq.(C.21) could be easily simplified and yields expression (8a) for the susceptibility in paraelectric phase.

From the strain field (C.16), (C.17) one could derive the following displacement field

$$u_1(x_1,x_2,x_3) = \left(u_0 + u_x\frac{x_3}{h}\right)x_1, \quad u_2(x_1,x_2,x_3) = \left(u_0 + u_x\frac{x_3}{h}\right)x_2,$$

$$u_3(x_1,x_2,x_3) = \int_{-h/2}^{x_3} u_{33}(\tilde{x}_3)d\tilde{x}_3 - u_x\frac{x_1^2 + x_2^2}{2h}. \tag{C.22}$$

Vertical displacement at the surface $x_3 = h/2$ is

$$u_3(x_1,x_2,h/2) = h\left(\frac{q_{11}(c_{11}+c_{12}) - 2c_{12}q_{12}}{(c_{11}^2 + c_{11}c_{12} - 2c_{12}^2)}\right)\langle\eta_3^2\rangle + \frac{12(f_{12}c_{11} - c_{12}f_{11})}{(c_{11}^2 + c_{12}c_{11} - 2c_{12}^2)}\left\langle\frac{\partial\eta_3}{\partial x_3}\frac{x_3}{h}\right\rangle\frac{x_1^2 + x_2^2}{2h} \tag{C.23}$$

Here the first term is independent on lateral coordinates; it is the manifestation of spontaneous strain of the free system (it could be also rewritten as $hQ_{11}\langle\eta_3^2\rangle$ and typically do not exceed several percents), while the second term corresponds to the particle bending induced by flexoeffect.

**Appendix D. Polarization distribution in ferroelectric nanowires**

A linearized solution for the polarization distribution and the averaged polarization was derived in Ref.[43] without the flexoelectric effect. For the considered case of paraelectric phase in Eq.(8) it acquires the form:

$$P_3(\rho) = \sum_{n=1}^{\infty} \frac{E_n^0 J_0(k_n \rho/R)}{a_R + g_{12}^*(k_n/R)^2} = \sum_{n=1}^{\infty} \frac{E_0 J_0(k_n \rho/R)}{a_R + g_{12}^*(k_n/R)^2} \frac{2J_1(k_n)E_0}{k_n(J_0^2(k_n) + J_1^2(k_n))} =$$
$$= \frac{E_0}{a_R}\left(1 - \frac{J_0(\rho/R_0)}{J_0(R/R_0) - (\lambda^*/R_c)J_1(R/R_0)}\right) \quad (D.1a)$$

$$\overline{P}_3 = \sum_{n=1}^{\infty} \frac{E_0}{a_R + g_{12}^*(k_n/R)^2} \frac{4J_1^2(k_n)}{k_n^2(J_0^2(k_n) + J_1^2(k_n))}$$
$$= \frac{E_0}{a_R}\left(1 - \frac{2}{R/R_0}\frac{J_1(R/R_0)}{J_0(R/R_0) - (\lambda^*/R_0)J_1(R/R_0)}\right) \quad (D.2a)$$

Where we used that, $R_0 = \sqrt{-g_{12}^*/a_1(T)}$, coefficients

$$E_n^0 = \frac{E_0}{M_n}\int_0^R \rho d\rho\, J_0\left(\frac{k_n \rho}{R}\right) = \frac{E_0}{M_n}\left(\frac{R^2}{k_n}J_1(k_n)\right) = \frac{2J_1(k_n)E_0}{k_n(J_0^2(k_n) + J_1^2(k_n))}$$ and the norm

$M_n = \int_0^R d\rho\, \rho\, J_0^2\left(\frac{k_n \rho}{R}\right) = \frac{R^2}{2}(J_0^2(k_n) + J_1^2(k_n))$. Here $J_0(k_n)$ and $J_1(k_n)$ are Bessel functions of the zero and first orders respectively. The roots $k_n$ depend over the ratio $(\lambda^*/R)$ in accordance with equation $J_0(k_n) - (\lambda^*/R)k_n J_1(k_n) = 0$.

The transcendental equation $a_R + g_{12}^*(k_n/R)^2 = 0$ for the determination of the transition temperature $T_{cr}(R)$ at a given radius $R$ as well as for the critical radius $R_{cr}(T)$ at a given temperature $T$ (that corresponds to the second order phase transition from ferroelectric to paraelectric phase) acquires the form:

$$J_0\left(R\sqrt{-\frac{a_R(T,R)}{g_{12}^*}}\right) - \lambda^*\sqrt{-\frac{a_R(T,R)}{g_{12}^*}} J_1\left(R\sqrt{-\frac{a_R(T,R)}{g_{12}^*}}\right) = 0. \quad (D.3)$$